\documentclass[pdflatex,sn-mathphys-num]{sn-jnl}

\usepackage{graphicx}%
\usepackage{multirow}%
\usepackage{amsmath,amssymb,amsfonts}%
\usepackage{amsthm}%
\usepackage{mathrsfs}%
\usepackage[title]{appendix}%
\usepackage{xcolor}%
\usepackage{textcomp}%
\usepackage{manyfoot}%
\usepackage{booktabs}%
\usepackage{algorithm}%
\usepackage{algorithmicx}%
\usepackage{algpseudocode}%
\usepackage{listings}%
\usepackage{multirow}
\usepackage{array}
\newcolumntype{C}[1]{>{\centering\arraybackslash}p{#1}}

\usepackage{lineno}

\usepackage{setspace}

\newtheoremstyle {defname}
{}{}
{\normalfont}
{}
{\bfseries}
{}
{.5em}
{\thmnote{#3}}

\theoremstyle{defname} 
\newtheorem*{namedprinciple}{}

\begin{document}

\title{Optimising Cryogenic Wiring for Microwave-frequency Millikelvin Quantum Processors}

\author{\fnm{Adrien} \sur{Di Lonardo$^{\textrm{1,2,}}$}}

\author{\fnm{Giorge} \sur{Gemisis$^{\textrm{1,2,}}$}}

\author{\fnm{Juan Pablo} \sur{Dehollain$^{\textrm{1,}}$}}

\author{\fnm{Nathan K.} \sur{Langford$^{\textrm{1,}\,\mathrm{\ast}}$}}

\affil[1]{
    \orgdiv{Centre for Quantum Software \& Information, School of Mathematical \& Physical Sciences}, 
    \orgname{University of Technology Sydney}, 
    \orgaddress{\city{Sydney}, \country{Australia}}}

\affil[2]{
    \orgname{Sydney Quantum Academy}, 
    \orgaddress{\city{Sydney}, \country{Australia}}}

    \abstract{
\begingroup
\renewcommand{\thefootnote}{}
\footnotetext{$^\ast$Corresponding author: nathan.langford@uts.edu.au.}
\addtocounter{footnote}{-1}
\endgroup
Cryogenic capacity is a key bottleneck in the scaling of superconducting quantum processors, critically impacted by control wiring. 
Currently, wiring architectures are often designed following a small set of ``best-practice'' heuristics, little changed since they emerged for early, small-scale systems, and not rigorously evaluated for wider contexts.
Here, we adopt a whole-system numerical modelling approach and present a systematic optimisation for the coaxial wiring design of a superconducting quantum processor.
We identify optimised configurations that substantively outperform conventional designs in device temperature, noise burden, qubit capacity and other practical metrics.
Importantly, our analysis delivers varied outcomes for different system contexts and operating regimes, illustrating the benefits of a systematic approach when there is no universal solution.
Our results evaluate and sometimes challenge conventional wisdom in relation to design factors such as the utility of 0dB attenuators.
We also use a wide exploration of cable configurations---directly enabled by our holistic and flexible numerical modelling approach---to develop more rigorously supported general principles for future cryogenic system design, based on a new conceptual framework to describe attenuator cascades that draws on analogies to low-noise amplifier chains.
Our approach, design outcomes, and generalised analyses should also be readily applicable to other cryogenic quantum computing platforms.
Finally, we support ready adoption of this approach through a web-based graphical tool which streamlines the analysis of the proposed framework under different contexts. 
Our results illustrate how our approach can help maximise the available computational resources of quantum processors at all system scales.
}




\maketitle


\section{Introduction}\label{sec:intro}

Many leading quantum computing platforms, such as those based around microwave-frequency, solid-state qubits, need to operate at cryogenic temperatures to suppress thermal excitations and relaxation in the quantum processor.
Although crucial for achieving high-fidelity coherent control, cryogenic operation creates two significant challenges that make it non-trivial to design the classical control layer responsible for the generation, delivery, and processing of control signals:
1) Control lines introduce ``passive'' thermal flux loads at each stage of the cryogenic system, which must be carefully balanced progressively against the finite cooling capacities available that diminish as stage temperature decreases; 
and 2) Control lines unavoidably open up channels to connect room-temperature thermal noise direct to the low-temperature qubits, and the attenuation required to adequately suppress this noise in turn creates active thermal loads due to signal power dissipation.
The fundamental tension between the number of individual signal lines required to control and measure a quantum processor, and their resulting heat and noise burden, upper bounds the number of qubits and qubit performance (and in turn computational complexity) the processor can support.
This tension must be carefully optimised across a complex system of interacting design choices to maximise a quantum processor's computational ``reach'', a task which will become increasingly challenging as processors push towards larger sizes.
Nevertheless, since quantum algorithm complexity scales exponentially with system size, even modest improvements in processor performance metrics such as error rates or operational qubit capacity can result in outsized increases in computational power.

Ultimately, profound technological innovations will likely be necessary to realise system scales that can deliver practical quantum advantage, and though they may not yet be realised at any practical scale, numerous technologies are being pursued, each with their own benefits and challenges.
Novel approaches include: 
\emph{photonic link architectures}, which aim to use microwave-optical signal conversion to interrupt the high thermal conductivity associated with metallic signal-carrying wires with low thermally conductive optical fibres, with pioneering demonstrations for AC single-qubit control only~\cite{Lecocq2021,Youssefi2021};
\emph{low-temperature electronics architectures based on cryogenic CMOS technologies}, which aim to locate control electronics as close as possible to the quantum processor in temperature to reduce room-temperature control and control latency and even monolithic control integration, with demonstrations usually limited to operation at $4\,\text{K}$~\cite{VanWinckel2022,Xue2021,Underwood2024} and millikelvin operation strongly limited by very low millikelvin cooling powers~\cite{Bartee2025};
and \emph{low-temperature electronics architectures based on single flux quantum (SFQ) logic}, which aim to drive qubit control operations using superconducting circuits that encode digital information in quantized magnetic-flux pulses, 
with demonstrations still limited by low-temperature cooling power budgets~\cite{McDermott2014,Leonard2019,Liu2023,Walter2026,Jordan2026}, and no demonstrations of parametric flux control or entanglement generation despite existing proposals \cite{Wang2023}.  

By contrast, conventional coaxial cabling architectures are still widely used for microwave frequency qubit control, deployed in quantum computing applications on devices with 10s to 100s of qubits~\cite{Mooney2021,Kim2023,Cao2023,Kam2024,Acharya2023,Gao2025}.
There are many degrees of freedom that determine this performance for a full coaxial cabling stack, including attenuator and filter configurations, cable selection, and thermalisation methods and locations.
Cable configuration selections are often based around standard ``best practice'' solutions---e.g., using $20$--$30\,\text{dB}$ attenuation on each of the 4K and mixing chamber (MXC) stages~
\cite{schuster2007phd,johnson2011phd,blumoff2017phd,ansmann2009phd,kelly2015phd,goeppl2009phd}---or simplistic design ``rules of thumb''---e.g., that low-temperature attenuation should match the relative low-temperature Johnson noise power~\cite{fink2010phd,bianchetti2010phd}.
A more comprehensive study analysed individual elements of a cryogenic wiring design for circuit QED, but compared performance for only a small number of standard configuration choices~\cite{Krinner2019}, while another measured cryogenic and electrical performance for a particular choice of coaxial cable material~\cite{Raicu2025}.
Given the prevalence of architectures using cryogenic coaxial cabling, and the relative immaturity of other alternative technologies, without a more systematic analysis and optimisation of cryogenic performance, it is still an important open question how to best design cryogenic wiring for large-scale cryogenic quantum processors. 

In this work, we address this problem by developing a holistic numerical modelling approach for optimising the performance of a full stack of coaxial cryogenic wiring.
In the model, we incorporate all physical processes which underpin the interaction between the cryogenic infrastructure and the control wiring, as understood from the best available literature, and use this model to systematically evaluate the space of design choices in the context of our own fixed-capacity BlueFors LD-400 fridge, and the signal requirements associated with the operation of our prototype quantum processors.
We demonstrate that, through systematic whole-system analysis, we can achieve substantial improvements in both performance and capacity relative to existing ``best practice'' designs and heuristics.
We show we can easily optimise designs for different constraints and custom performance requirements, such as maximising cable flexibilty for device prototyping, with cost functions generally comprising competing performance metrics and optimal design choices hence often depending in turn on application-specific design priorities.
Our analysis arrives at various design choices which differ from conventional ``best practice'' based on quantitative performance evaluations of a holistic design and specific system contexts.
We also generalise our analysis, however, to formulate a new design principle for optimising noise suppression in attenuation configurations, by analogy with principles of noise propagation in amplifier chains.
Finally, to streamline this process, we have also developed a web-based GUI tool which makes it easy for others to analyse and optimise performance for their own experimental contexts.
Overall, our results highlight the value of systematic whole-system analysis to optimise performance and capacity, and thus meaningfully improve the accessible computational resources, at any scale.

Although this work focusses on the design of coaxial cabling for microwave control of superconducting quantum circuits, the general approach is agnostic to the specific cryogenic quantum platform or quantum computing paradigm.
For example, it should be relatively straightforward to extend the heat dissipation and noise propagation routines used in our numerical model to accommodate the heat dissipation mechanisms and noise transmission/generation characteristics of other types of control lines, such as the microwave Electron Spin Resonance (ESR) lines \cite{Dehollain2012,Veldhorst2015}, DC voltage lines in twisted-pair wiring looms, and baseband pulsing lines \cite{Li2018, GonzalezZalba2021, Boter2022, Kunne2024} commonly used for spin-qubit quantum computing platforms.

The remainder of the paper is organised as follows:
In Section~\ref{sec:modelling}, we outline the key details of the numerical model we use to evaluate performance for a full coaxial wiring stack.
We then illustrate our systematic approach in Section~\ref{sec:optimisation}, and the impact it has on design decisions, by using our model to evaluate the coaxial cabling design space for prototype quantum processors in our BlueFors LD-400 fridge;
we also discuss the implications of conventional and unconventional design choices.
In Section~\ref{sec:different_constraints}, we demonstrate the flexibility of our approach by re-analysing the design for different design priorities, and the value of holistic modelling by optimising design directly for maximising qubit capacity, comparing against naive implementations based on conventional designs and existing heuristics.
In Section~\ref{sec:new_heuristic}, we formulate a new design principle via analogy with noise in amplifier chains, and demonstrate how it compares to our optimal results and existing, more simplistic heuristics.
Finally, in Section~\ref{sec:conclusion}, we summarise and compare our approaches against ``best practice'' designs, consider the limitations of a priori numerical modelling, and discuss our web-based GUI tool---``Cryowala''---for streamlining customised systematic system optimisation.

\section{Details of the ``Cryowala'' Numerical Model}\label{sec:modelling}

Most of the physics associated with the heat loads and noise propagation in cryogenic wiring setups has been long and well established, deriving mostly from textbook physics concepts, e.g., about blackbody radiation, thermal conductivity and electronics.
A detailed overview of the main physics involved is provided already in \cite{Krinner2019}.
The most complex elements are arguably related to the detailed cryo-engineered response of a dilution refrigerator, but we avoid this complexity in the conventional way by capturing this response through phenomenological cooling-power curves, derived either empirically or via a simple heuristic model (again see \cite{Krinner2019}).

We now briefly outline key details of our model, following conventions adopted in \cite{Krinner2019}.

\subsection{Systematic Holistic Optimisation}

Despite the simplicity of the physics involved, as discussed above, cryogenic microwave cabling designs have commonly been based on convention and design ``rules of thumb'' rather than systematic analysis and optimisation~\cite{Krinner2019, Raicu2025}.
For many years, this was not a significant limit to growth in quantum processor sizes, since cryogenic performance requirements were outpaced by growth in the cryogenic capacity of dilution refrigerator technology.
More recently, recognising such growth could not continue indefinitely, new design benchmarks were set by \cite{Krinner2019}, which studied performance requirements for a large-scale commercial dilution refrigerator wired for superconducting quantum circuit systems, with performance optimised for each kind of cryogenic control line individually and overall performance studied only for a small number of discrete configurations.

Our alternative approach was to consider the full cryogenic wiring stack holistically, building a flexible full-system model of heat loads and noise, which allows systematic optimisation---including of individual components---based on overall cryogenic performance.
This involves: simulating heat loads from thermal conduction (passive loads) and signal dissipation (active loads) for all coaxial lines, with different attenuator configurations chosen for each line type; using these heat loads to model the temperature response of the fridge stages; and using these temperatures to model the propagation of noise through the attenuators thermally anchored at each temperature stage.

Because key performance metrics like device temperature, noise temperature and qubit capacity are often competing, design decisions often involve inherent trade-offs, and there will rarely be an obvious unique cost function to define optimal performance.
Here, taking a holistic approach makes it relatively straightforward to systematically explore and develop targeted cabling solutions around bespoke, application-specific design constraints:
When trying to maximise the capacity of a particular cryogenic system for large-scale quantum processors, a genuinely optimised design for targeted requirements can provide substantive increases to control capacity, and consequently also computational power.
We demonstrate our systematic approach in a similar context to \cite{Krinner2019}, of coaxial cabling for microwave control of superconducting quantum circuits, but we note that the approach itself is far more generic.
Whatever technology a cryogenic system is using to provide precision quantum control, when that system is being designed to operate at the limit of its capacity (whether that be in terms of performance or volume/complexity, or a combination of both), it will likely always be necessary to optimise the system both systematically and holistically.

\subsection{Noise Propagation}
Qubit control cables necessarily connect eventually to the quantum processor, creating a channel for room-temperature thermal noise that must be strongly suppressed to achieve robust operation of microwave-frequency quantum devices.
Any resistive electrical component generates Johnson-Nyquist (thermal) noise, that increases with temperature, with the temperature-dependent power spectral density (assuming a matched load)~\cite{pozar2012microwave}:
\begin{equation}
    S_\text{JN}(T, f)=hf\,n_\text{BE}(T, f)=\frac{hf}{\exp{\left(\frac{hf}{k_\text{B}T}\right)}-1},
    \label{eqn:psd_JN}
\end{equation}
where $n_\text{BE}(T, f)$---the Bose-Einstein distribution in terms of
Planck's constant $h=6.626 \times 10^{-34} \, \text{J}\cdot\text{s}$ and Boltzmann's constant $k_\text{B}=1.38 \times 10^{-23} \, \text{J}\cdot\text{K}^{-1}$---effectively represents the photon spectral density in $\text{photons}\,\text{s}^{-1} \text{Hz}^{-1}$.
The power spectral density reduces to the classical Johnson-Nyquist noise, $S_\text{JN}(T, f)\approx k_\text{B}T$ in the low-frequency limit, where $hf \ll k_\text{B}T$, and is exponentially suppressed in the high-frequency limit, where $hf \gg k_\text{B}T$.

Low-temperature attenuation suppresses incoming noise, including thermal noise, from electronics and components at higher temperature stages (including from room temperature electronics), at the cost of introducing additional lower-temperature Johnson-Nyquist noise at their own temperature, $T_\text{att}$.  Given an attenuation rating, $A$, the total output noise from an attenuator is:
\begin{equation}
    S_\text{out}= \frac{S_\text{in}}{A_\text{att}} + \left(1 - \frac{1}{A_\text{att}}\right) S_\text{JN}(T_\text{att})
    \label{eqn:attenuator_noise}
\end{equation}
In this way, with sufficient attenuation, attenuators can effectively ``thermalise'' the transmitted noise (near) to the noise baseline for their own temperature.

\subsection{Passive loads}
Coaxial control cables also introduce heat loads to the cryogenic system by directly thermally linking hotter and colder stages.
The thermal gradient between any two thermal anchoring points results in a heat flux through the coaxial cable which is determined by Fourier's law,
\begin{equation}
    P = \frac{A}{L}\int_{T_1}^{T_2}\sigma(T)dT
    \label{eqn:fouriers_law}
\end{equation}
depending on the temperature-dependent thermal conductivity, $\sigma(T)$ of the cable materials, cross-sectional area $A$ and length $L$.
As the coaxial cable is composed of a concentric inner conductor and outer shield separated by a dielectric, the total heat flux is the sum of the heat flux through each material, $P_\text{total} = P_\text{inner} + P_\text{dielectric} + P_\text{outer}$.

The coaxial cables are thermally anchored to each stage so as to progressively thermalise them without overloading the finite cooling capacity of the millikelvin stage where the device is located.
The dielectric, however, thermally insulates the inner conductor from the outer, preventing ready thermalisation of the inner conductor to each stage via the direct physical contact which is used to thermalise the outer conductor.
If, however, a cable can be anchored to the stage through an attenuator, that attenuator provides the direct connection the inner conductor needs for thermalisaton, in addition to providing attenuation of the transmitted signal.
Our modelling assumes that the inner conductor is only thermalised to stages where an attenuator or filter is present to provide the required thermal link between inner and outer conductors.

\subsection{Active loads}
Control signals also introduce cryogenic heat loads, produced by dissipation along the signal path:

\textbf{Attenuator loss.} The (resistive) attenuators used to reduce noise also directly absorb a $(1 - 1/A_\text{att})$ fraction of any incoming signal power (see Eq.~\ref{eqn:attenuator_noise}), and dissipate it as heat transferred to the temperature stage where the attenuator is thermally anchored:
\begin{equation}
    P_\text{active}=\left(1 - \frac{1}{A_\text{att}}\right)\,P_\text{in}
    \label{eqn:P_att}
\end{equation}

\textbf{High-frequency cable loss.} Coaxial cables also exhibit intrinsic transmission loss, typically specified by manufacturers as a total loss rate in $\text{dB}\,\text{m}^{-1}$ that includes both conductive and dielectric losses.
Our model assumes that, for typical materials and microwave frequencies, this loss arises primarily from conductive losses associated with the flow of carrier currents through finite-resistivity inner and outer conductors, concentrated within the skin depth, $\Delta(f)$.
Since the currents associated with travelling fields are split between the inner and outer conductors, so is the dissipated power:
\begin{align}
    P_\text{loss}=\left(1-\frac{1}{A_\text{loss}}\right)P_\text{in}=P_\text{loss,inner}+P_\text{loss,outer}.
    \label{eqn:P_loss}
\end{align}
Moreover, since the currents are split equally, the power dissipation is split between inner and outer conductive loss in relation to the cross-sectional area within a skin depth of each surface, which is approximately $2\pi r\,\times\,\Delta(f)$, because the microwave-frequency skin depth is much less than the conductor radial dimensions.
The fraction of power dissipated in each surface is therefore:
\begin{align}
    \frac{P_\text{loss,inner}}{P_\text{loss}}
    &=\frac{\rho_\text{inner} r_\text{outer}}{\rho_\text{outer} r_\text{inner}+\rho_\text{inner} r_\text{outer}}
    =\frac{r_\text{outer}}{r_\text{inner}+r_\text{outer}}, \\
    \frac{P_\text{loss,outer}}{P_\text{loss}}
    &=\frac{\rho_\text{outer} r_\text{inner}}{\rho_\text{outer} r_\text{inner}+\rho_\text{inner} r_\text{outer}}
    =\frac{r_\text{inner}}{r_\text{inner}+r_\text{outer}},
    \label{eqn:P_cable_microwave}
\end{align}
where $r_\text{outer}$ is the radius of the outer conductor's inner surface, where the current flows (not the outer radius of the coaxial cable), and the second equality in each line holds for the common case where inner and outer conductors are made from the same material (and hence where $\rho_\text{inner}=\rho_\text{outer}$).

As with attenuator loss, the heat dissipated due to cable loss is transferred to the fridge at thermal anchoring points.
Here, we assume a worst-case scenario where power losses from any cable section are dissipated at the next coldest thermal anchoring point: i.e., the next temperature stage for heat dissipated in the outer conductor, and the next attenuator or filter for heat dissipated in the inner conductor.

\textbf{Low-frequency cable loss.} Finite-resistivity coax inner and outer conductors also give rise to ohmic losses for DC flux biasing currents, according to $P=I^2R$.
As for AC currents, we assume that conduction losses dominate over dielectric losses at DC, but contrastingly, DC currents flow through the inner conductor only.
We again assume that the dielectric effectively thermally insulates the inner coax conductor from the outer shield, so that heat generated by DC currents is transferred to the fridge at thermal anchoring points.

For flux lines, the final attenuator for DC input currents is generally on the 4K plate, so above that stage, we can assume that DC-generated heat is dissipated at the next attenuator.
Below that, it gets more complicated.
While it is fairly straightforward to model how much total heat is generated due to ohmic losses along the length of any given coaxial cable segment, it can be very difficult to predict from first principles exactly where that heat is dissipated, since DC currents commonly sink to the fridge ground through very complex pathways.

Due to measurements illustrating that material specifications for cable resistances do not accurately predict the actual losses, Ref.~\cite{Krinner2019} instead employed an empirical Ohmic model to predict how heat generated below the 4K plate is dissipated on the mixing chamber and cold-plate stages.
(Note: Ref.~\cite{Krinner2019} measured negligible heat dissipation for the Still plate, likely because heat dissipation would be dominated there by externally applied heating needed to maximise operational mixing-chamber cooling powers.)
We adopt the same approach, including, for the purposes of this analysis, starting with the same underlying effective resistances for the CP and MXC stages (as measured empirically for 2.19mm stainless-steel cables in \cite{Krinner2019}), and then scaling the resistive heat generation for other cable types according to their dimensions and material properties:
\begin{equation}
    \frac{R_\text{cable}}{R_\text{emp.}} = \frac{\rho_\text{cable}}{\rho_\text{emp.}} \frac{L_\text{cable}}{L_\text{emp.}} \frac{A_\text{emp.}}{A_\text{cable}}
    \label{eqn:cable_flux_scaling}
\end{equation}

\subsection{Temperature Response Modelling}
The temperature of the fridge stages are dynamic and depend on the heat loads applied to them.
This is because each stage has some effective temperature-dependent cooling process, where the rate of heat flux removed increases quadratically with increasing temperature.
The instantaneous temperature profile of the fridge stages is then given by the temperatures at which the incoming heat fluxes---from both the coaxial lines and any other thermal transfer processes through the fridge structure---are in equilibrium with the cooling power of the each stage.
Not surprisingly, it is both prohibitively and unnecessarily difficult to model all the minute details of the fridge cooling processes from first principles, due to the substantial engineering complexity of state-of-the-art cryogenic systems.
Nevertheless, the temperature profile can be characterised empirically by measuring equilibrium temperatures for each stage as a function of known, manually applied heat loads.
This is then used to predict the temperature response of each fridge stage from the heat flux modelled from the coaxial lines.
The captured data and regression modelling for the temperature response of our BlueFors LD-400 fridge is discussed further in Appendix~\ref{app:temp_response}.

\begin{table}[h]
\caption{Cooling power budgets for each stage of the BlueFors LD-400 fridge
}\label{tab:cooling_budget}
\begin{tabular*}{\textwidth}{@{}lll@{}}
\toprule
Refrigerator Stage & Target Maximum Operating Temperature (K) & Available Cooling Power(W)\\
\midrule
50K & 46.2 & 5 \\
4K & 3.86 & 350$\times 10^{-3}$ \\
Still & 1.24 & 30$\times 10^{-3}$	\\
CP & 130$\times 10^{-3}$ & 300$\times 10^{-6}$ \\
MXC & 21$\times 10^{-3}$ & 20$\times 10^{-6}$ \\
\botrule
\end{tabular*}
\end{table}

Throughout this paper, to contextualise the simulated heat loads, we express them as a fraction of some assigned cooling budget for each stage given in Table~\ref{tab:cooling_budget}.
These cooling budgets essentially describe the cooling power available at the target maximum operating temperatures for each stage.

\section{Systematic Optimisation of Coaxial Lines for Performance}\label{sec:optimisation}

In this section, we describe a systematic evaulation and optimisation for the coaxial wiring stack supporting prototype quantum processors in our BlueFors LD-400 fridge. 
We used a fixed total line capacity of 32 lines, supporting a 14-qubit device with 14 microwave drive lines for single qubit control, 14 microwave flux biasing lines for parametric control, and 2 pairs of readout lines consisting of input and output lines to the readout resonators.
At the device, we expect to operate microwave drive signals for single qubit control at -75 dBm, with an approximate duty cycle of 33\%.
We assume that half of the transmon qubits are operated at their top sweet spot ($\sim 0\,\Phi_0$), and that half are asymmetric transmon qubits operated at their bottom sweet spot ($\sim 0.5\,\Phi_0$), which translates to a flux biasing current of $\sim0\,\text{mA}$ and $\sim2\,\text{mA}$ respectively using an average mutual inductance of $\sim0.25\,\Phi_0/\text{mA}$.

Throughout this paper, unless otherwise stated, whichever lines are not the subject of a particular analysis are included with fixed, default values, as described in Appendix.~\ref{app:default-configurations} and Table~\ref{tab:default-configurations}.

\subsection{Microwave Drive Line Attenuation}\label{subsec:ac}

\begin{figure}[t]
	\centering
 	\includegraphics[width=\linewidth]{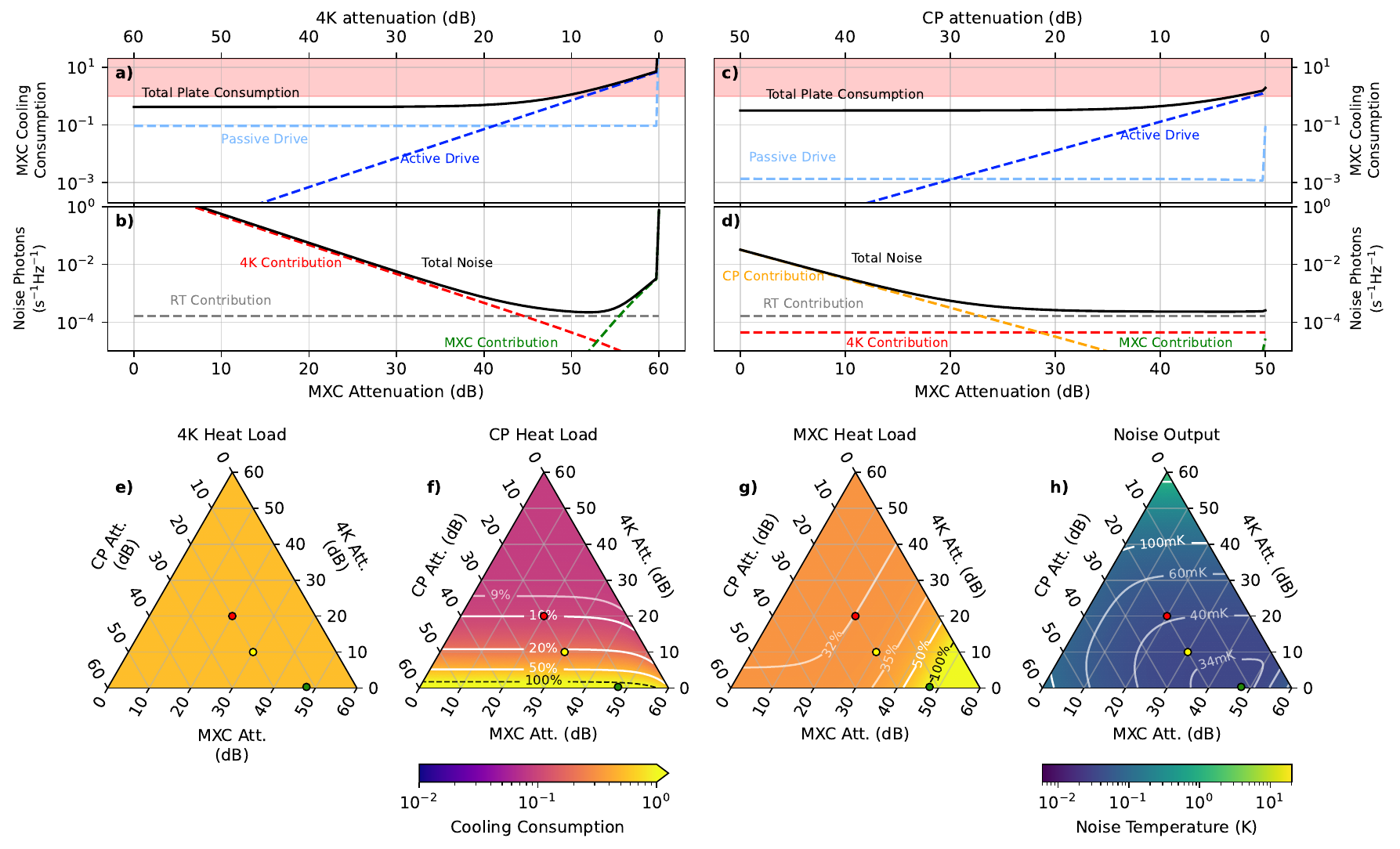}
 	\caption{
        Systematic analysis of microwave drive line attenuation configurations for fourteen (14) 2.19mm stainless-steel coaxial drive lines (219-SS-SS) (with flux and output lines fixed to defaults, see Tab.~\ref{tab:default-configurations}).
        Total attenuation is fixed at $60\,\text{dB}$, with output power normalised at $-72\,\text{dBm}$ with a duty cycle of 33\%.
        Expected heat loads (a) and noise performance (b) at the MXC stage where the device is located (at the output of the microwave lines) are shown for configurations with attenuation at the 4K and MXC stages only.
        Panel (a) shows both total full-system heat loads, and passive and active heat load contributions, expressed as a fraction of the MXC cooling power budget (see text).
        Panel (b) also shows total noise performance, and individual room-temperature and stage contributions, expressed in terms of thermal noise photon spectral density.
        At the noise minimum, at the $10$-$50\,\text{dB}$ split, the MXC stage is already overloaded, with temperature beginning to increase dramatically. 
        Panels (c) and (d) show heat loads and noise for improved configurations achieved by splitting the $50\,\text{dB}$ low-temperature attenuation between the CP and MXC stages.
        Within the large plateau of minimum noise, we identify as optimal the configuration consuming the least cooling power, given by $10$-$20$-$30\,\text{dB}$ for the 4K-CP-MXC stages.
        The flat dominant noise contribution from the RT electronics is explored further in Section~\ref{sec:new_heuristic}.
        Panels (e--g) and (h) show, as ternary plots, heat loads and noise, respectively, for full configuration space analysis of 4K-CP-MXC attenuation splittings, to assess the full performance landscape.
        Our selected configuration (yellow point) achieves near-optimal noise performance, better than best practice designs~\cite{Krinner2019} (red point), as well as a configuration from generalised attenuation configuration analysis in section~\ref{sec:new_heuristic} (green point).
    }
    \label{fig:ACLines}
\end{figure}

The full configuration space of attenuator configurations is four-dimensional, allowing for attenuators to be thermally anchored to all but one of the five fridge stages, assuming no attenuators anchored to the Still plate:
This is a common design choice to avoid subjecting the Still to fluctuating active heat loads from fluctuating input signals, which may drive undesirable temperature fluctuations at the mixing chamber, due to the direct relationship between Still temperature, the helium evaporation and circulation rate, and the cooling power of the dilution unit at the mixing chamber.
To further simplify the analysis here, we also fix the total attenuation to $60\,\text{dB}$---borrowing from standard practice---thus reducing the configuration space to three dimensions:
It requires around $70\,\text{dB}$ of attenuation (e.g., $60\,\text{dB}$ of attenuation plus another $10\,\text{dB}$ from direct cable loss) to attenuate room-temperature thermal noise to roughly the same---at $6\,\text{GHz}$---as the thermal noise produced by a $50\,\Omega$ component at $\sim30\,\text{mK}$.
From there, while any additional attenuation further reduces the room temperature noise, it also increases the input signal strengths required for qubit drives, which in turn increases the active loads across all fridge stages.
This can make the challenge of balancing noise suppression and thermal flux more challenging.

Due to the competing drivers imposed by noise requirements, cooling power limitations and quantum device capacity, as well as application-specific priorities, there is no clear, universal cost function on which cryogenic performance can be objectively and unambiguously optimised.
This makes a brute force optimisation across the full three- or four-dimensional configuration space unhelpful, as well as being impractical.
Furthermore, we ideally want to develop useful guiding principles, and validate (or invalidate) existing cryo-microwave engineering heuristics and rules of thumb.
Again, brute-force optimisation is not very instructive for elucidating the relationship between individual design choices and performance.
To arrive at, and build confidence in, a more reliable result, we instead undertake a systematic investigation by sequentially exploring targeted configuration subspaces that are more tractable to visualise and interpret.
The process, and key results, are illustrated in Figure~\ref{fig:ACLines}.

We first investigate the 4K-MXC subspace in Figure~\ref{fig:ACLines}~(a, b), for which all other attenuators are fixed to 0dB.
This resembles the design space commonly considered in early circuit QED literature where attenuation was only employed at the 4K and MXC stages~\cite{schuster2007phd,johnson2011phd,ansmann2009phd,goeppl2009phd}.
We plot the heat loads on the MXC stage (a) and the total noise at the output of the microwave drive lines (b) to illustrate the trade-off between noise suppression and thermal flux which results in a noise minimum arising at around the $10\,\text{dB}$-$50\,\text{dB}$ configuration.
To elucidate this noise minimum, it is useful to plot—--alongside the total output noise---the individual noise contributions from room temperature electronics and different attenuator states, based on Equation~\ref{eqn:attenuator_noise}.
This shows that the minimum occurs where the noise is dominated by the flat room temperature noise floor, with attenuator contributions both substantially lower.
Away from the minimum, the total noise is dominated by either insufficient attenuation of the 4K noise contribution, or a sharply increasing MXC noise contribution from increasing MXC temperatures arising because high MXC active loads exceed the available cooling capacity%
\footnote{A discontinuity appears in the heat load and noise curves when the 4K attenuation becomes $0\,\text{dB}$: this arises because it is assumed that $0\,\text{dB}$ attenuation means that no attenuator is installed, removing the physical connection which thermally anchors the inner conductor of the coaxial cable to the 4K stage.
Consequently, a greater passive load in the drive lines is suddenly transmitted to the MXC stage, causing a spike in heat and noise.
}.
The precise location of this minimum is therefore sensitive to the detailed balance of MXC heat loads against cooling powers across all heating contributions from all signal and cable types.

Note that, the MXC passive load is dynamically responsive to the change the change in MXC plate temperature as the active load far exceeds the cooling capacity, however, the magnitude of this change is small giving the appearance of a flat response.
This is because the temperature gradient of relevance here is that between the 4K and MXC stages---since the inner cable conductor is not thermalised to the fridge stages between these plates ---which is large compared to the change in MXC temperature.

Earlier ``best practice'' designs often employed a $20\,\text{dB}$ attenuator on the 4K plate~\cite{johnson2011phd,blumoff2017phd,kelly2015phd,goeppl2009phd}, since $4\,\text{K} / 300\,\text{K} \sim 10^{-2}$, which is not aligned with the location of the noise minimum observed in Fig.~\ref{fig:ACLines}~(b).
But while the optimal configuration achieves better noise in this scenario than the configuration with $20\,\text{dB}$ at the 4K plate, it does so at the cost of significantly increased cooling consumption (and hence device temperature and qubit capacity).
We can achieve the best of both worlds, however—--noise performance similar to the $10\,\text{dB}$-$50\,\text{dB}$ configuration with MXC cooling consumption even lower than the $20\,\text{dB}$-$40\,\text{dB}$ configuration---by fixing $10\,\text{dB}$ on the 4K plate and splitting the remaining $50\,\text{dB}$ attenuation between the CP and MXC stages.
As shown in Figure~\ref{fig:ACLines}~(c, d), this creates a large plateau of near-optimal noise performance for a range of CP-MXC configurations, and we can now select a configuration which consumes near-minimal MXC cooling power: giving $10$-$20$-$30\,\text{dB}$ attenuation for the 4K-CP-MXC stages.
This selection also deviates from the ``best practice'' design identified in \cite{Krinner2019}, which argued an alternative standard configuration of $20$-$20$-$20\,\text{dB}$ was optimal~\cite{fink2010phd,bianchetti2010phd}.

Having identified these basic behaviours through more focussed analyses, we can validate our choices in a wider context, by looking at the full two-dimensional configuration space of 4K-CP-MXC attenuation splittings in Figure~\ref{fig:ACLines}~(e--h), with the 50K attenuation assumed to be zero%
\footnote{We also explored configurations involving 50K attenuation, see Appendix~\ref{app:50K}, but find that it is does not provide substantial benefit to either the heat or noise performance of the microwave drive lines in this context.}.
In this full configuration space, we validate that our optimised configuration delivers near-optimal noise performance at the expense of 
a marginally higher cooling consumption than the $20$-$20$-$20\,\text{dB}$ configuration, for both MXC and CP stages.
This is a worthwhile trade-off as it still operates well within the total available cooling budget of our fixed-capacity BlueFors LD-400 fridge, 
and improved control line noise will improve qubit coherence times and gate fidelities.

Compared with naive implementation of existing designs and heuristics, our systematic evaluation of the design space has 
enabled us make meaningful design decisions which achieve a better balance between noise suppression and thermal flux.
This analysis also highlights that noise minima occur when the room temperature noise contribution is dominant.
We discuss the implications of this observation for establishing new design heuristics in Section~\ref{sec:new_heuristic}.

\subsection{Flux Biasing Lines}\label{subsec:dc}

\begin{figure}[t]
	\centering
 	\includegraphics[width=\linewidth]{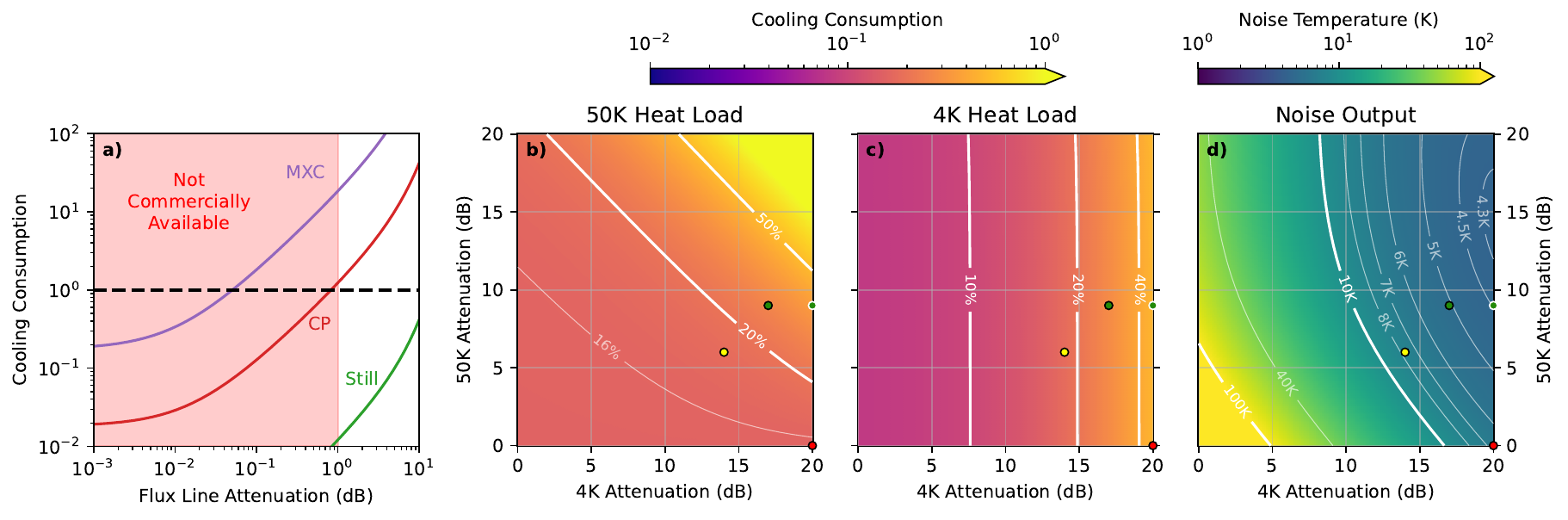}
 	\caption{
        Systematic analysis of ``DC'' flux-biasing line attenuation configurations for fourteen (14) 2.19mm stainless-steel flux lines (219-SS-SS), with half the lines at each of $0$ and $2\,\text{mA}$ biasing currents fixed at the device (with flux and output lines fixed to defaults, see Tab.~\ref{tab:default-configurations}).
        Expected total flux line heat loads on the Still, CP and MXC plates (a) are shown for marginal attenuation values, indicating that attenuation below the 4K stage is of limited practical use, even at attenuation values below what is commercially available.
        This has implications for the utility of $0\,\text{dB}$ attenuators (see Figure~\ref{fig:0dB}).
        Panels (b,c) and (d) show heat loads and noise, respectively, for the full remaining configuration space across the 50K and 4K plates up to $20\,\text{dB}$ on each plate.
        Here, noise performance is expressed as the effective noise temperature of the output spectral density.
        Comparisons between our design (yellow point) and existing best practice~\cite{Krinner2019} (red point) are shown, as well as configurations from the generalised attenuation analysis in section~\ref{sec:new_heuristic} (green points).
        Some configurations (black-bordered points) are used in performance comparisons in \ref{fig:performance_comparison}.
        We identify an optimised 6dB-14dB configuration, which achieves a comparable noise temperature to best-practice designs while operating further within the available cooling capacity of the 4K plate.
    }
    \label{fig:DCLines}
\end{figure}

As the flux biasing currents directly control the frequency of superconducting qubits, noise current in the flux biasing lines is a source of qubit dephasing.
However, the relative sensitivity of the qubits to noise current in the flux biasing lines is lower than that in the microwave drive lines.
The weaker coupling of noise current to the qubits results in 
flux biasing signal power (based on $2\,\text{mA}$ flux biasing current) 
which is approximately seven orders of magnitude higher than the microwave drive powers used for our devices, resulting in higher signal dissipation.
These large signal powers largely prohibit attenuation of flux biasing lines below the 4K stage as even marginal dissipation of the signal power can overwhelm the cooling budgets for the MXC and CP stages, as shown in Figure~\ref{fig:DCLines}~(a).
We explore the consequences of this for the use of 0dB attenuators in flux biasing lines in Section~\ref{subsec:0dB}.
The weaker coupling of flux biasing lines to the qubits also means that noise suppression can be less stringent.
Upper bound estimates of decoherence time as a result of noise in the flux lines are estimated to be $T_2^* \approx 46\,\mu\text{s}$, $426\,\mu\text{s}$ and $2333\,\mu\text{s}$ \cite{Krinner2019} when the flux lines use $0\,\text{dB}$, $10\,\text{dB}$ and $20\,\text{dB}$ of attenuation on the 4K plate, respectively.
On this basis, thermalising flux biasing line noise to the temperature of the 4K plate should ensure that noise current in the flux-biasing line is not a limiting factor for qubit coherence.
This regime of operation where the sensitivity of the qubits to the control lines is reduced is analogous to applying more attenuation to the control lines
and further illustrates the trade-off between noise suppression and cooling consumption which makes cryogenic wiring design non-trivial.

In practical terms, the design space of attenuator configurations is restricted to only the 50K and 4K plates, which we fully explore in Figure~\ref{fig:DCLines}~(a-c).
In constrast to the widely deployed $0\,\text{dB}$-$20\,\text{dB}$ configuration on the 50K and 4K plates respectively, we find that employing $6\,\text{dB}$-$14\,\text{dB}$ attenuators achieves a largely comparable noise temperature while reducing the total 4K cooling consumption by a substantial factor of 3.
This further illustrates the value of systematic evaluation of the design space to meaningful design choices 
which achieve substantial improvements in performance.

\subsection{0dB Attenuators}\label{subsec:0dB}

\begin{figure}[t]
	\centering
 	\includegraphics[width=0.9\linewidth]{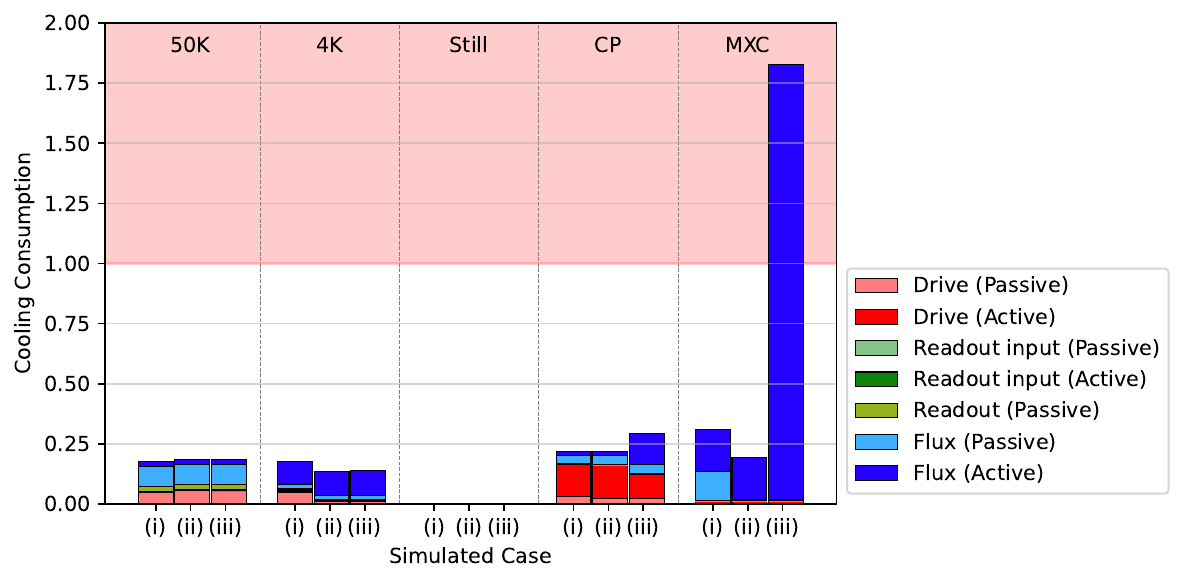}
 	\caption{
        Practical limitations of $0\,\text{dB}$ attenuation.
        Numerical modelling of heat loads is shown comparing the optimised configurations identified from Figures~\ref{fig:ACLines} and~\ref{fig:DCLines} with (i) no $0\,\text{dB}$ attenuators, (ii) ideal lossless $0\,\text{dB}$ attenuators, and (iii) lossy $A_\text{real} = 0.1\,\text{dB}$ attenuators.
        When included, $0\,\text{dB}$ attenuators are placed at all locations with no existing attenuation.
        The $0\,\text{dB}$ attenuators provide thermalisation of the inner coaxial conductor which reduces passive loads on subsequent (lower) plates, and active drive loads on the CP.
        However, even a marginal loss of $0.1\,\text{dB}$ (better than standard commercially available performance) introduces untenable active flux loads on the MXC plate where the device is located.
    }
    \label{fig:0dB}
\end{figure}

Here, we discuss operational considerations in relation to the use of $0\,\text{dB}$ attenuators that are revealed by our systematic analysis. 
These are components which are specifically designed to address the challenge of inner conductor thermalisation without 
generating additional signal power dissipation. 
This is significant because it in principle allows for progressive thermalisation of the inner cable at all stages where thermal anchoring was previously unavailable,
including at the Still plate and below the 4K plate for the flux lines.
This would reduce the passive heat load burden on the lower plates.

Recent work has characterised the real-world effectiveness of the thermalisation provided by commercially available $0\,\text{dB}$ attenuators across a range of cryogenic temperatures~\cite{Klomp2026}.
In practice, it is also important to consider the impacts of non-zero microwave dissipation in real-world $0\,\text{dB}$ attenuators, arising from unwanted electrical resistance and dielectric losses.
Standard commercial $0\,\text{dB}$ attenuators are likely to have losses up to $0.2$--$0.3\,\text{dB}$, or even more~\cite{xma-0dB,quantummicrowave-0dB}.
Importantly, because of the high signal power of flux biasing lines, 
even minor attenuation can result in heat dissipation that represents a substantial fraction of available cooling powers on the lower plates.
Even at $0.1\,\text{dB}$, the active flux dissipation could be completely unmanageable for the mixing chamber, as shown in Figure~\ref{fig:DCLines} (a) and Figure~\ref{fig:0dB}.

We compare the performance of the fully optimised configurations from Sections~\ref{subsec:ac} and~\ref{subsec:dc} 
with the addition of $0\,\text{dB}$ attenuators at all previously unattenuated locations in Figure~\ref{fig:0dB}.
We simulate the cryogenic wiring under both the assumptions of $A_\text{real}=0\,\text{dB}$ loss and $A_\text{real}=0.1\,\text{dB}$ loss. 
Both the lossy and lossless scenarios do indeed provide substantial improvements in the passive loads, reducing the 4K AC line
passive load by 75\% and the MXC DC line passive load by 98.6\%, which can create useful gains in line capacity or plate temperatures.
However, the introduction of even small losses massively increases the DC line active load, to 1.83 times the MXC cooling budget.
This doesn't fully preclude the use of $0\,\text{dB}$ attenuators in cryogenic wiring, but highlights that these components cannot be employed 
blindly and should be carefully considered in the design process.
For example, while standard $0\,\text{dB}$ attenuators on the cold plate could be manageable in terms of heat load, and the pay-off may be low:
In our target configurations, the passive load from flux lines was not a performance-limiting factor, and the presence of $0\,\text{dB}$ attenuators at the cold plate does not reduce the active loads on MXC plate.

\subsection{Cable Materials}\label{subsec:cable}

\begin{figure}[t]
	\centering
 	\includegraphics[width=\linewidth]{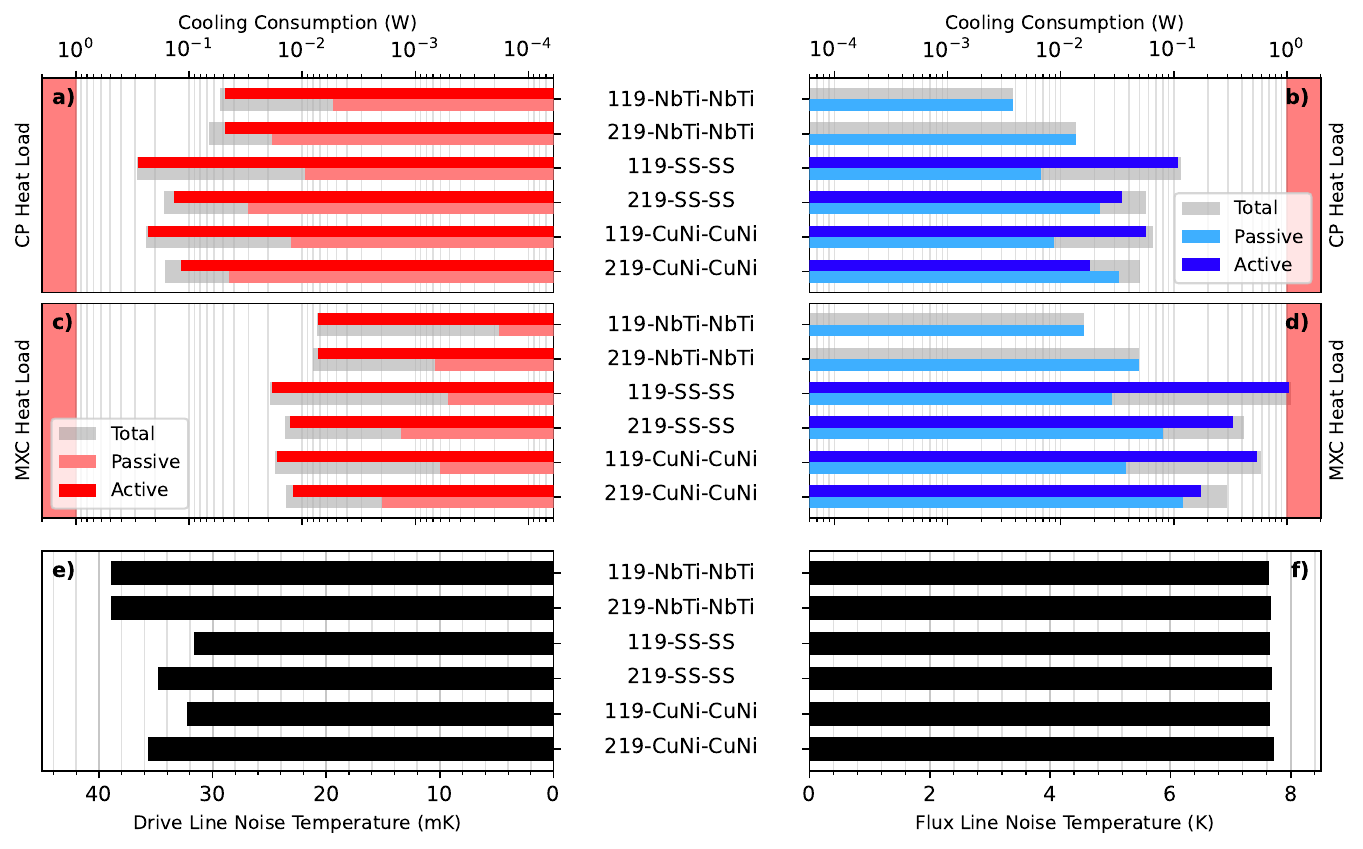}
 	\caption{
        Comparing the impacts of cable materials and sizes on optimised performance.
        Modelling includes 14 drive and 14 flux biasing lines with optimised configurations identified in Figures~\ref{fig:ACLines} and~\ref{fig:DCLines}.
        Heat loads are shown for the CP (a, b) and MXC plates (c, d) divided by passive and active loads including attenuator losses and cable losses, for both drive lines (a, c) and flux biasing lines (b, d).
        Active loads and passive loads are inversely related, as cable resistivity and thermal conductivity are inversely related.
        Drive (e) and flux (f) line noise performance is shown as effective output noise temperature.
        Based on these results, we choose 2.19mm CuNi cables for the flux biasing lines, against established best practice, to mitigate the dominant active flux loads on the MXC, and 2.19mm SS cables for the drive lines, to balance heat loads, noise suppression, and financial cost (not shown).
        Superconducting NbTi cables below the 4K are included for comparison.
        They provide the lowest passive and active loads, which is promising for managing heat loads in large scale deployments, and would achieve equivalent noise suppression with additional fixed attenuation installed.
        They are, however, currently cost-prohibitive, and therefore generally only used for output lines.
    }
    \label{fig:CableMaterial}
\end{figure}

The selection of cable type plays an important role in the performance, as they directly set the passive loads and provide additional noise suppression.
Using the attenuator configurations selected from the Sections~\ref{subsec:ac} and~\ref{subsec:dc}, 
we compare the total noise suppression and cooling consumption on the CP and MXC using a range of cable materials and sizes for the 
drive and flux biasing lines in Figure~\ref{fig:CableMaterial}.
Clear trends emerge relating to the physical properties of the cable materials.
Passive loads increase with thermal conductivity (stainless-steel, SS, to copper-nickel, CuNi) and cross-sectional area (1.19mm to 2.19mm outer diameter),
while actively loads increase with resistivity inversely to passive loads.
Active flux line loads resulting from the ohmic dissipation of the flux biasing currents are the dominant contribution 
to the total cooling consumption on the MXC stage.
There is little difference in flux line noise performance between the cable types as no additional cable loss is provided
at DC frequencies.
We subsequently select 2.19mm CuNi cables for the flux biasing lines to minimise the MXC heat loads and plate temperature, 
despite the higher passive loads and reduced noise suppression.
This design choice, which emerged from whole-system analysis, is also in contrast to existing rules of thumb which target low-thermal conductivity cables 
to minimise passive heat loads \cite{Krinner2019}.
With respect to the drive lines, simulatenous consideration of the heat and noise performance allows us to select 2.19mm SS cables 
which deliver the best balance of the trade-off between heat loads and noise suppression.

We also simulate the performance of superconducting NbTi cables below the 4K stage for comparison to quantify the purported benefits
in the literature \cite{Krinner2019}, as they are attractive for their low thermal conductivity and zero cable losses.
Indeed, they provide more than a factor of 2 lower passive loads for the same diameter and exhibit zero ohmic dissipation
which reduces the total flux line loads by up to an order of magnitude.
This is extremely promising for managing heat loads in large scale deployments, in particular for device architectures with a high number of flux lines,
for example, devices which employ tunable couplers~\cite{Arute2019QS}.
It should be noted, however, that the superconducting cables require additional attenuation to achieve the same noise suppression as lossy cables.
These cables also present a financial barrier to large-scale adoption as they are typically 5 times more expensive other cables simulated here.

\section{Systematic Optimisation Under Different Constraints}\label{sec:different_constraints}

\subsection{Optimising for Prototyping Flexibility}\label{subsec:transformable}

\begin{figure}[t]
	\centering
 	\includegraphics[width=0.65\linewidth]{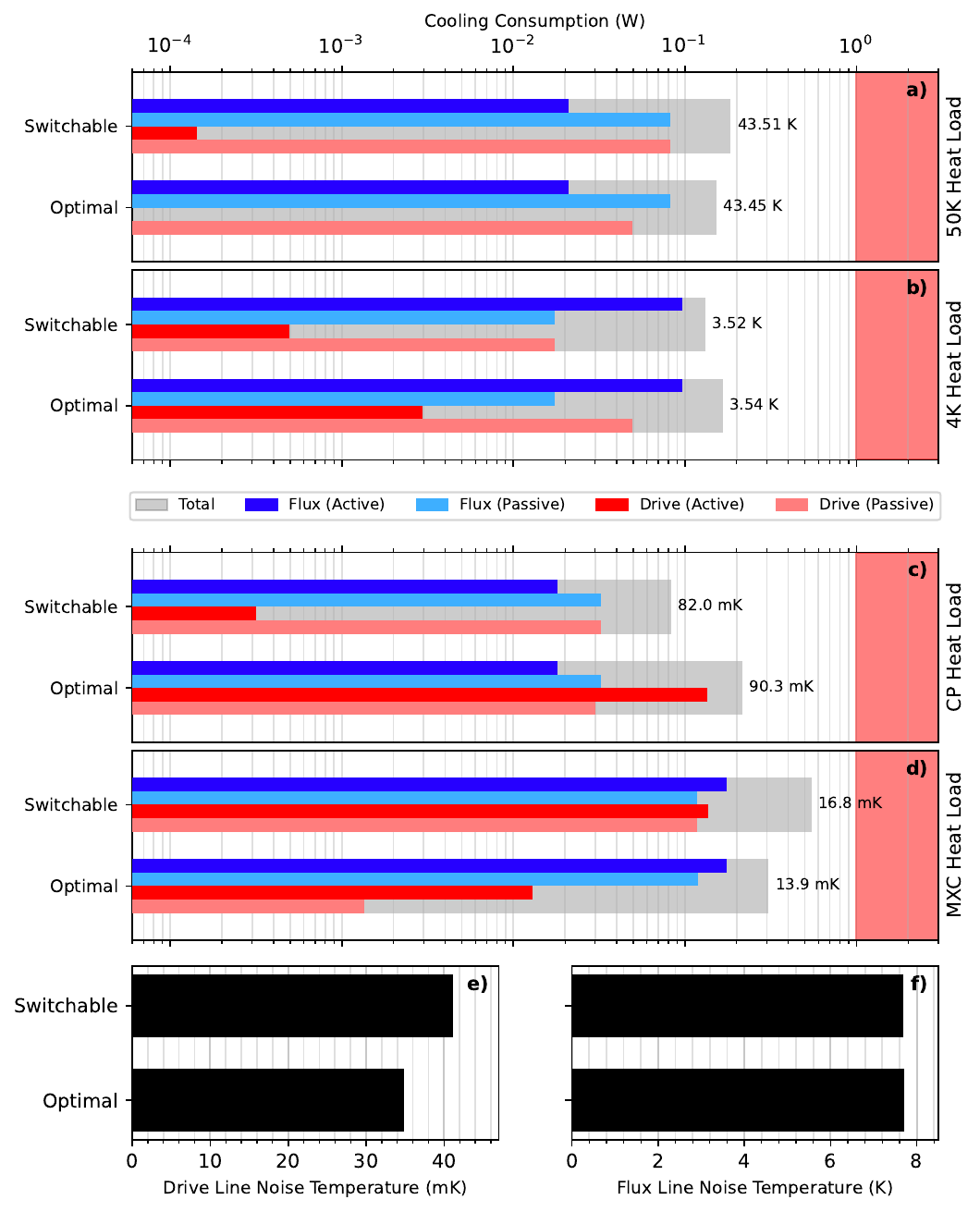}
 	\caption{
        Evaluation of a ``switchable'' line design (see text), which can be readily converted to either microwave drive or ``DC'' flux biasing signals between cooldowns, without modifying the basic wiring stack.
        The ``Optimal'' configuration incorporates the optimised selections from from Figures~\ref{fig:ACLines},~\ref{fig:DCLines}, and~\ref{fig:CableMaterial}, and is compared against a fully ``switchable'' installation of 28 ``switchable'' lines, with 14 used for microwave drives, and 14 for ``DC'' flux biasing.
        Heat load comparisons are shown across the 50K (a), 4K (b), CP (c), and MXC (d) plates, along with estimated temperature responses.
        Heat loads from the readout-in and -out lines are included in the total heat loads, but not shown directly as their performance is identical in both designs.
        Noise performance is shown for the drive lines (e) and flux lines (f) as effective output noise temperature.
        Decreased performance of the switchable design in MXC temperature and drive line noise represents a trade-off against the additional flexibility provided by the design (see text).
        In Fig.~\ref{fig:performance_comparison}, we include, for comparison, a configuration including half of the 28 drive and flux lines installed as switchable lines, achieving a compromise in performance between the two cases shown above, for a degree of flexibility.
    }
    \label{fig:TransformableLines}
\end{figure}

Aside from the main constraints of plate temperatures, noise temperatures and line capacities, there are also other practical 
considerations which influence design decisions. 
For example, financial considerations inform the decision to use superconducting cables, despite their universal performance benefits. 
For our prototype devices, a single fridge will be used to rotate through multiple devices quickly with different wiring requirements, 
which means that wiring complexity and re-installation time is an important factor.
For many cryogenic architectures, most cables are installed in dense bundles packed in small volumes, meaning that removing a single cable 
may require complete disassembly of a cable array. 
Inevitably, there are costs associated with disassembly and re-installation times, and additional risks of component damage or compromising replicability of cryogenic wiring performance.
As such, it is desirable to have some degree of flexibility where some lines can accommodate different signal input types (e.g., both microwave and flux biasing signals) without needing to change the physical wiring.

Here, the drive and flux biasing lines have vastly different requirements for noise suppression and cooling consumption, which makes it impossible to implement a single static line which is suitable for both.
Instead, building on early designs~\cite{schuster2007phd,johnson2011phd,blumoff2017phd,ansmann2009phd,kelly2015phd}, we assume the wiring stack is fixed above the MXC stage but allow for components to be swapped out underneath, since changing attenuator configurations in this location doesn't impact commissioned semi-rigid coaxial cable installations.
We therefore aim to design a ``switchable'' cable which uses a modified version of the optimised flux biasing lines, $[A_\text{50K}=6, A_\text{4K}=14, A_\text{Still}=0, A_\text{CP}=0]\,\text{dB}$, above the MXC stage.
The cable then can then be configured to support microwave drive signals by adding a $A_\text{MXC}=40\,\text{dB}$ attenuator on the MXC stage.
A 2.19mm CuNi cable is chosen to be suitable for both as a flux biasing and microwave drive line.

We simulate a full wiring architecture with 28 switchable lines and 4 readout lines in Figure~\ref{fig:TransformableLines} to exaggerate the performance differences 
with the original optimised design from Section~\ref{sec:optimisation}.
Out of the 28 switchable lines, 14 are configured as drive lines and 14 as flux biasing lines.
The switchable design overall delivers $~3.2\,\text{mK}$ higher MXC temperature and $~6\,\text{mK}$ higher drive line noise temperature, owing to a combination
of the increased passive loads and lower loss from the CuNi cables and the suboptimal arrangement of drive line attenuation which can be understood
from previous analysis in Figure~\ref{fig:ACLines}.
Overall, these trade-offs are within operational targets and acceptable given the flexibility provided by the design. 
In practice, choose only a fraction of total lines to be switchable to achieve a sufficient level of flexibility for the full cryogenic wiring stack.
This allows for dedicated optimised lines to still be used.
This also means that the expected cooling consumption and device temperature will be some value between the two extremes of the switchable and dedicated designs.
When we compare overall fridge performance in Figure~\ref{fig:performance_comparison}, we elect to install half of the control lines as switchable lines.

\subsection{Optimising for Qubit Capacity}\label{subsec:capacity}

\begin{figure}[t]
	\centering
 	\includegraphics[width=\linewidth]{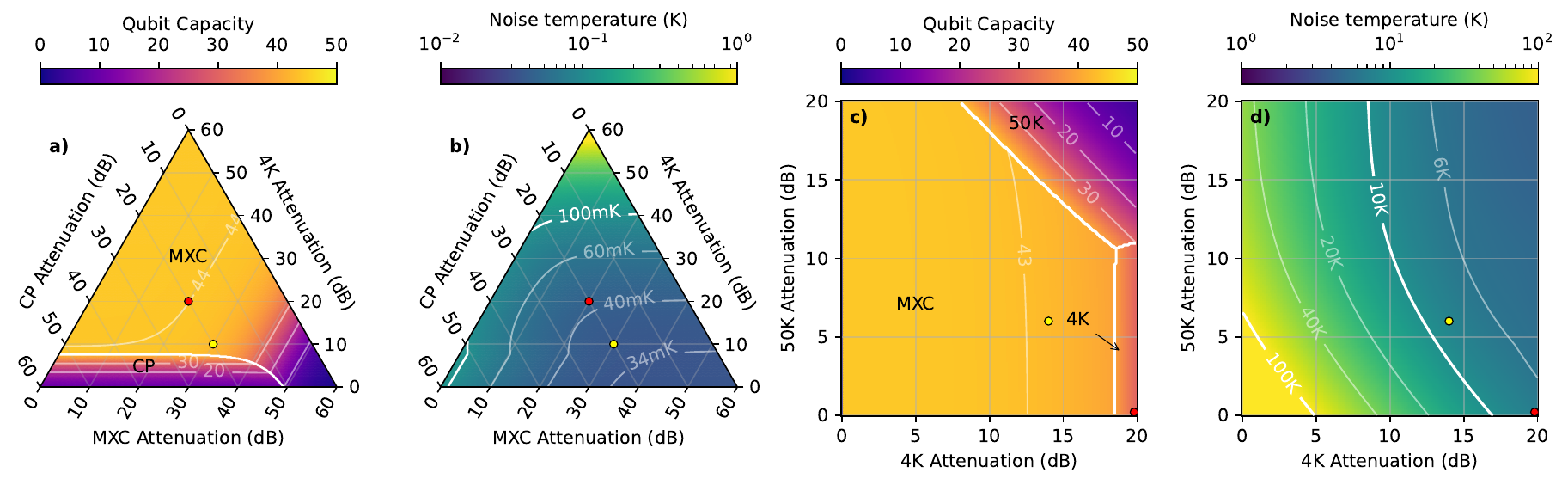}
 	\caption{
        Qubit control capacity estimation under different drive (a, b) and flux line (c, d) attenuator configurations for our BlueFors LD-400 fridge.
        We assume a ratio of 1 drive, 1 flux, and 0.125 readout lines per qubit's worth of control capacity, and ignore practical spatial constraints relevant for installation.
        Total capacity (a, c) is plotted as the maximum number of qubits which can be operated within the available cooling capacity of all stages.
        A bottleneck stage is identified as that which has the highest cooling consumption per qubit and regions for each bottleneck are shown.
        Plots of noise temperature (b, d) demonstrate the trade-off between cooling consumption and noise suppression.
        Our design (blue point), amalgamated from the analysis in Section~\ref{sec:optimisation}, provides substantial improvement in qubit capacity in noise performance over ``best practice'' designs (red point).
    }
    \label{fig:QubitCapacity}
\end{figure}

Qubit and line capacity is a critical metric for realising large-scale quantum processors.
Here, we pivot our existing analyses to demonstrate that a systematic approach can directly optimise the qubit capacity,
Crucially, this is only possible through holistic numerical modelling in which the whole-system performance is considered, 
as the qubit capacity is a function of all heat loads and is sensitive to interactions between them.
To estimate qubit capacity, we assume a fixed ratio of 1 drive line and 1 flux biasing line per qubit and 1 multiplexed readout line pair per 8 qubits, which is typical for superconducting qubit devices.
We omit the additional flux biasing lines that might be required for tunable coupler devices for this specific analysis%
\footnote{Devices with tunable couplers would have at least 1 more flux line per qubit (on average) and would be more sensitive to flux-line heat load contributions.} to provide a fair comparison with best practice designs in other literature \cite{Krinner2019}.
The above ratios allow for an estimate of cooling consumption per qubit on each stage.
The total capacity is then given by the number qubits which can be supported on the stage with the highest fractional cooling consumption per qubit, as this stage will bottleneck the overall system.

Following the configuration space analyses from Figures~\ref{fig:ACLines} and~\ref{fig:DCLines}, 
we explore the qubit capacity across a range of attenuation configurations for the drive and flux lines in Figure~\ref{fig:QubitCapacity}.
We divide the configuration space into regions based on which stage is the bottleneck for qubit capacity, which is a convenient way to understand how to deliver further gains.
The optimised designs from Section~\ref{sec:optimisation} deliver substantial improvements in the overall balance of qubit capacity and noise performance over ``best practice'' designs.
These designs are near-optimal considering that traversing in any direction in the configuration space will either significantly increase the noise temperature or reduce the qubit capacity.

\section{Attenuation Design Heuristics}\label{sec:new_heuristic}

Choosing attenuator configurations for coaxial cryogenic wiring involves two key elements: 1) choosing what total attenuation to use, and 
2) choosing how to distribute that attenuation amongst the different temperature stages.
At the simplest level, in relation to the former, the more attenuation is used, the less noise reaches the quantum device, but since the necessary control signal power is fixed by the device design irrespective of the cable configuration, more attenuation also means higher input signal powers and more active head load dissipation.
In relation to the latter, the more the attenuation distribution is skewed towards the low-temperature stages nearest the device, the lower the noise that reaches the quantum device, but the more heat is dissipated where there is least cooling power; on the other hand, while skewing the attenuation towards higher-temperature stages exploits larger cooling powers, this leaves less low-temperature attenuation to block the resulting high-temperature thermal noise.

These competing demands obviously create trade-offs that need to be weighed and balanced in the design process.
Standard practice is to first decide what total attenuation is required (or possible) while remaining manageable, and then to decide how much attenuation to put on each stage, placing a significant emphasis on keeping heat loads and stage temperatures low, especially for the final MXC plate.
Thus, where possible, this generally involves prioritising more attenuation on the higher stages that have more cooling power.
This is essentially the approach followed in~\cite{Krinner2019}, for example, where a particular total attenuation of $60\,\text{dB}$ was chosen to achieve a specified $10^{-3}$ thermal noise photon occupation at the device.
For the systematic optimisation described in the first part of this paper, we targeted the same total attenuation, to ensure our designs achieved a minimum standard of performance and facilitate direct comparison with the conclusions from~\cite{Krinner2019}, but conducted a wider exploration of attenuation distributions.
So far, we also primarily focussed on empirical modelling to optimise cryogenic cable designs, focussing on attenuator configurations, but we now aim to identify more generally applicable guiding principles for attenuator design, to cross-check the outcomes we obtained from direct optimisation, and compare those principles with existing heuristics and standard practices.

\subsection{The ``Equal Balance'' Principle}

Conventional wisdom recommends following an ``equal balance'' design heuristic to achieve optimal noise performance without excess active heat dissipation, with attenuation at a given stage being chosen so that the attenuated higher-temperature noise from the preceding stage is roughly equal to the added Johnson-Nyquist noise generated by the attenuator at the current stage (according to Eq.~\ref{eqn:attenuator_noise}).
The discussion in~\cite{Krinner2019} correspondingly recommends reference attenuation values of $A_i = S_\text{JN}(T_{i{-}1}, f)/S_\text{JN}(T_{i}, f)$ for stage $i$, which defines the saturation balance point from Eq.~\ref{eqn:attenuator_noise} where the transmitted and added noise components are equal.
In situations where the standard Johnson noise formula holds (moderate temperatures or low frequencies)---that is, $hf \ll k_\text{B}T$, such that $S_\text{JN}(T, f)\approx k_\text{B}T$---the ``equal balance'' reference attenuation simplifies to $A_i \approx T_{i{-}1}/T_{i}$.
While this is the formula which is often used in practice and is a good approximation for a dilution refrigerator's higher-temperature stages, the equal balance condition depends very sensitively on frequency for operating frequencies above the thermal noise cut-off.

For a modest dilution fridge base temperature of, say, $30\,\text{mK}$,  low-frequency noise calculations predict around $40\,\text{dB}$ total attenuation is required to match the room-temperature noise contribution, which is consistent with the $40$--$50\,\text{dB}$ aggregate attenuation typically chosen for early circuit QED experiments~
\cite{schuster2007phd,johnson2011phd,blumoff2017phd,ansmann2009phd,kelly2015phd,goeppl2009phd}.
For typical circuit QED operating frequencies of 4--$8\,\text{GHz}$, however, the low-frequency assumption is no longer valid for the lower temperature stages, with cut-off frequencies ($hf_{\rm c}=k_\text{B}T$) of around 440 MHz and $2.7\,\text{GHz}$ for the MXC and CP stages, respectively (at the target temperatures given in Table~\ref{tab:cooling_budget}).
At $6\,\text{GHz}$, around $70\,\text{dB}$ total attenuation is required to balance the room-temperature noise contribution to the high-frequency Johnson-Nyquist noise emitted by a $30\,\text{mK}$ MXC attenuator, consistent with a choice of total $60\,\text{dB}$ from attenuators (plus roughly $10\,\text{dB}$ more attenuation from cable loss).
This already increases to $90\,\text{dB}$, however, to balance room-temperature noise to a $21\,\text{mK}$ MXC attenuator at $6\,\text{GHz}$.

These arguments also apply to the distribution of attenuation.
For example, early conventional designs generally split the attenuation between two stages (4K and MXC) only~
\cite{schuster2007phd,johnson2011phd,blumoff2017phd,ansmann2009phd,kelly2015phd,goeppl2009phd}, and employed $20\,\text{dB}\approx S_\text{JN}(300\,\text{K}, f)/S_\text{JN}(4\,\text{K}, f)\approx 300\,\text{K}/4\,\text{K}$ of attenuation on the 4K plate~\cite{johnson2011phd,blumoff2017phd,kelly2015phd,goeppl2009phd}.
While the low-frequency Johnson-Nyquist formula would then predict that around $20\,\text{dB}$ is needed to thermalise 4K noise down to $40\,\text{mK}$, the full frequency-dependent calculation shows that in fact more than $40\,\text{dB} \approx S_\text{JN}(4\,\text{K}, f)/S_\text{JN}(40\,\text{mK}, f)$ is required.
Note, however, that over $70\,\text{dB}\approx S_\text{JN}(4\,\text{K}, f=6\,\text{GHz})/S_\text{JN}(21\,\text{mK}, f=6\,\text{GHz})$ of total attenuation would be required to reach a base temperature of $\sim21\,\text{mK}$ at the MXC stage.
This would in turn result in $10^3$ times more signal dissipation on the MXC stage to achieve the same drive line signal power at the device.
As noted in~\cite{Krinner2019}, there is little opportunity to offset this by increasing attenuation at the 4K stage, due to the diminishing returns brought by increasing attenuation once the noise floor of that stage is reached.

The discussion in~\cite{Krinner2019} concludes that it is not practical to exploit the full noise floor available at a 20 mK MXC stage, due to cooling power limitations, taking the perspective that it is important to keep active heat loads on low-temperature stages far below their available cooling powers.
For typical total attenuations of $50$--$60\,\text{dB}$, it is therefore common for circuit QED experiments to operate with less attenuation than is consistent with genuinely equal noise balance between all stages.
At $60\,\text{dB}$, even given an additional $8$--$10\,\text{dB}$ of line attenuation, that would only ``thermalise'' the room-temperature noise contribution down to around $32$--$33\,\text{mK}$.
Furthermore, in terms of distributing the attenuation, the general approach is therefore to start at the top of the fridge, install just enough attenuation at each intermediate stage to ``thermalise'' the noise coming from the previous stage, before installing the remainder from a target attenuation budget at the mixing chamber.
Interestingly, the optimal configurations we identified in the first part of this paper (e.g., in Sec.~\ref{subsec:ac}) do \emph{not} appear to conform very closely to either the equal balance principle or this standard approach:
for example, an ``equal balance'' heuristic suggests that at least $18$--$19\,\text{dB}$ should be used at the 4K plate, but our analysis predicts that optimal noise performance occurs instead at around $10\,\text{dB}$, with much higher attenuations installed on the lowest temperature stages (see Fig.~\ref{fig:ACLines}).

\subsection{Analysis of an Attenuator Cascade}

To explore the principles of attenuation design more rigorously, we start by applying Eq.~\ref{eqn:attenuator_noise} to a cascade of attenuators, to write down the total Johnson-Nyquist noise contribution after the final attenuator (as seen by the quantum device):
\begin{align}\nonumber
S_{\rm DUT}(f) &= \overline{S}_\text{RT} + \overline{S}_\text{4K} + \overline{S}_\text{CP} + \overline{S}_\text{MXC}
\\ \nonumber
&= \frac{S_\text{JN}(T_{\rm RT}, f)}{A^\prime_{\rm 50K}\bar{A}_{\rm 4K}\bar{A}_{\rm CP}\bar{A}_{\rm MXC}} \\ \label{eqn:attenuation_output_noise}
+ &\frac{S_\text{JN}(T_{\rm 4K}, f)}{\bar{A}_{\rm CP}\bar{A}_{\rm MXC}}\left(1{-}\bar{A}_{\rm 4K}^{-1}\right)
+ \frac{S_\text{JN}(T_{\rm CP}, f)}{\bar{A}_{\rm MXC}}\left(1{-}\bar{A}_{\rm CP}^{-1}\right)
+ S_\text{JN}(T_{\rm MXC}, f)\left(1{-}\bar{A}_{\rm MXC}^{-1}\right),
\end{align}
where $\overline{S}_i$ refers to the noise contribution at the device from stage $i$, $S_\text{JN}$ is calculated according to Eq.~\ref{eqn:psd_JN}, and $\bar{A}_j = A_j \ldotp A^\prime_j$ is the total attenuation at each stage from both attenuator ($A_j$) and intrinsic cable attenuation ($A^\prime_j$). 
In our modelling, we assume that cable attenuation generates thermal noise at the temperature of the cable's low-temperature side, which is why $\bar{A}$ is used in the $1-1/\bar{A}$ terms at each stage, instead of including line attenuation noise separately.
This is a slightly ``best-case'' assumption in terms of noise output, since the cables sustain a temperature gradient between their ends, but we expect this discrepancy to be minimal, since the cable attenuations are in any case small.
To simplify the above equation, we assumed that attenuators are installed only on the 4K, CP and MXC plates (in line with Fig.~\ref{fig:ACLines}).
From this equation, we can see that there are effectively three qualitatively different types of contributors to the noise seen at the device:
\begin{enumerate}
    \item First term: The room-temperature noise contribution sets a configuration-independent noise floor that depends only on total line attenuation.
    \item Last term: The contribution from the final temperature stage sets an overall noise floor that is also (approximately) independent of configuration: while it does depend on the attenuation at that final stage, it typically does so only very weakly, provided at least a modest amount of attenuation is installed there; this is necessarily the case for all low-noise configurations.
    \item Middle terms: The intermediate stages share features with both room-temperature and MXC terms, contributing in a similar fashion: 1) like the room-temperature contribution, their noise contributions depend significantly on the total attenuation installed at lower stages; and 2) like the MXC contribution, the thermal noise generated at each temperature stage depends nonlinearly on its own attenuation through the $(1-1/A)$ term.
    In contrast, however, these stages do not set any kind of configuration-independent noise floor.
    For example, their noise contributions can usually be reduced significantly below the first and last stages (in principle, even if not in practice).
    Furthermore, the noise contributions for the first and last stages do not depend on how attenuation is specifically distributed between the intermediate stages.
    It is therefore only the total noise contributed by the intermediate stages (not how it is split between them) that is important.
\end{enumerate}
The ratio between noise contributions from sequential stages (for intermediate and final stages, with minor modifications for the RT stage) is given by an expression which depends only on the underlying Johnson-Nyquist noise distributions at each temperature, and the attenuations installed each stage (and for typical attenuations, only weakly on the first-stage attenuator):
\begin{align}\label{eqn:noise-ratio-adjacent}
\frac{\overline{S}_i}{\overline{S}_{i-1}} = \frac{S_\text{JN}(T_{i}, f)}{S_\text{JN}(T_{i-1}, f)}\frac{\left(A_i-1\right)}{\left(1-A_{i-1}^{-1}\right)},
\end{align}
For a full ``equal-balance'' configuration, where all stages contribute equally, the total Johnson-Nyquist noise seen by the device (output noise) is obviously $n+1$ times larger than the absolute noise floor set by the final temperature stage.

More generally, however, the above descriptions and formula make explicit the two key parameter regimes alluded to in the previous section: the MXC-dominant regime, where the total line attenuation reduces the room-temperature noise contribution well below the noise floor set by the (final) MXC temperature stage; and the RT-dominant regime commonly realised in practical circuit QED experiments, where the (incoming) room-temperature noise remains well above the expected MXC noise, because of design constraints for total attenuation.
In the latter regime, increasing total attenuation will (approximately) proportionately decrease the noise seen by the device, up to the point where the room-temperature noise contribution balances the final-stage noise floor.
Beyond that point, the noise reduction quickly saturates, but the increased attenuation continues to linearly increase the active heat loads that need to be dissipated on the cryostat from control signals.
On the other hand, because the control signal power required at the device is independent of line attenuation, the active heat load on the MXC plate also depends only (but here linearly) on the MXC attenuation.

As noted above, in contrast to the first and last stages, there is no particular difference between the intermediate stages in terms of the way they contribute to the total device noise.
To help identify general design principles that do not depend on complex system-specific optimisations, we therefore reduce the number of free parameters under consideration by assuming the intermediate stages all contribute the same noise.
Under this assumption, the total device noise can be characterised by two independent ratios: $r_{{\rm JN},\,{\rm MXC}}$ and $r_{{\rm JN},\,i \neq {\rm MXC}}$, defined relative to the room-temperature noise contribution according to $r_{{\rm JN},i} = \overline{S}_i/\overline{S}_{\rm RT}$, giving:
\begin{align}\label{eqn:attenuation_output_noise_ratios}
S_{\rm DUT}(f) &= \overline{S}_{\rm RT}(T_{i}, f) \, \big[ 1 + (n-1) \, r_{{\rm JN},\,i \neq {\rm MXC}} + r_{{\rm JN},\,{\rm MXC}} \big].
\end{align}

We obtain further insight by drawing an analogy between the design of attenuator configurations and the design of low-noise amplifier chains.
It is well known that well-designed cascaded systems of amplifiers aim to ensure that added noise is dominated by the first amplifier in the chain~\cite{pozar2012microwave}.
The absolute noise performance of an amplifier chain can be described by its effective noise temperature, given by the Friis equation:
\begin{align}\label{eqn:friis}
    T_{\rm eff} &= T_1 + \frac{T_2}{G_1} + \frac{T_3}{G_1 G_2} + \ldots + \frac{T_n}{G_1 \cdots G_{n-1}} \\ \label{eqn:friis-mod}
    &= T_1 \left(1 + \frac{T_2}{T_1 G_1} + \frac{T_3}{T_1 G_1 G_2} + \ldots + \frac{T_n}{T_1 G_1 \cdots G_{n-1}}\right),
\end{align}
where $T_E$ is the total effective input-referenced noise temperature of the chain, and where $T_j$ and $G_j$ are the input-referenced noise temperature and gain of the $j^{\rm th}$ amplifier, respectively.
This equation shows both how the first amplifier sets a gain-independent noise floor to effective overall performance, and how this noise floor can be saturated in practice (satisfying the general design principle described above), by ensuring that the ratio of noise temperatures between any individual amplifier and the first, is (much) less than the total gain preceding that amplifier~\cite{pozar2012microwave}.
The Friis equation is strikingly similar to Eq.~\ref{eqn:attenuation_output_noise} for the total Johnson-Nyquist noise seen by the device in our attenuator cascade, but with the roles of input, output and components effectively reversed;
and identical re-written in terms of the total noise emitted at the output of the attenuator, $N_n = (1-1/A_n)S_\text{JN}(T_n, f)$ (from Eq.~\ref{eqn:attenuator_noise}), it becomes:
\begin{align}
N_{\rm DUT}(f) &= \frac{N_{\rm RT}(f)}{A^\prime_{\rm 50K}\bar{A}_{\rm 4K}\bar{A}_{\rm CP}\bar{A}_{\rm MXC}}
+ \frac{N_{\rm 4K}(f)}{\bar{A}_{\rm CP}\bar{A}_{\rm MXC}}
+ \frac{N_{\rm CP}(f)}{\bar{A}_{\rm MXC}}
+ N_{\rm MXC}(f),
\end{align}
Note, while amplifier noise temperatures are typically defined via the low-frequency Johnson-Nyquist noise formula $S_\text{JN}(T) = k_\text{B}T$, there is no reason why Eq.~\ref{eqn:friis} cannot also be defined in terms of full Boltzmann-distributed (frequency-dependent) noise, $T_{\rm eff}(f) = S_\text{JN}(T, f)/k_\text{B}$.

The attenuator cascade acts like an amplifier chain in reverse because of two conceptual differences: 1) because the attenuators are components with less than unity gain ($G_n = A_n^{-1} < 1$), and 2) because the relevant signal level is fixed at the output of the attenuator cascade and at the input of the amplifier chain.
In the latter, the relevant noise performance of an amplifier chain in any given context is defined by comparison with the intrinsic strength and signal-to-noise ratio of the input signal, hence why all noise terms in Eq.~\ref{eqn:friis} are referenced back to the input of the first amplifier in the chain.
In the former, however, the relevant signal strength required is fixed, in an absolute sense, at the output of the attenuator cascade (by the device design).
\emph{Its} relevant noise performance is therefore also defined by comparison with the signal strength required at the \emph{output} of the cascade.
So gain maps to attenuation, and input-referenced amplifier noise maps to output-referenced attenuator noise.

Taking this analogy at face value, general design principles for attenuator cascades should also be inverted from the well-established design principles for low-noise amplifier chains:
that is, for minimum noise, control-line configurations should aim to ensure that the output noise is dominated by the noise contribution from the \emph{final} attenuator (the MXC-dominant regime described above), instead of the first.
In other words, the design goal should be to choose the total attenuation to be significantly larger than the reference value identified by a na\"ive ``equal balance'' design heuristic~\cite{Krinner2019}---i.e., $r_{{\rm JN},\,n}\gg1$ (and similarly for the added noise emitted by attenuators at all other stages, i.e., $r_{{\rm JN},\,n} \gg r_{{\rm JN},\,i \neq n}$).
Crucially, however, these design principle can also be impacted strongly and directly by the fridge's full cooling-power and heat-load response. 
In the rest of this section, we explore the impact of fridge response on general attenuation design principles using our full, holistic fridge model.
Our aim is not to focus on the specific details of our particular fridge response, but rather on how including a fridge response affects how general design principles should be applied.

\subsection{Guiding Principles for Attenuator Design: Flux Lines}

\begin{figure}[!t]
	\centering
 	\includegraphics[width=\linewidth]{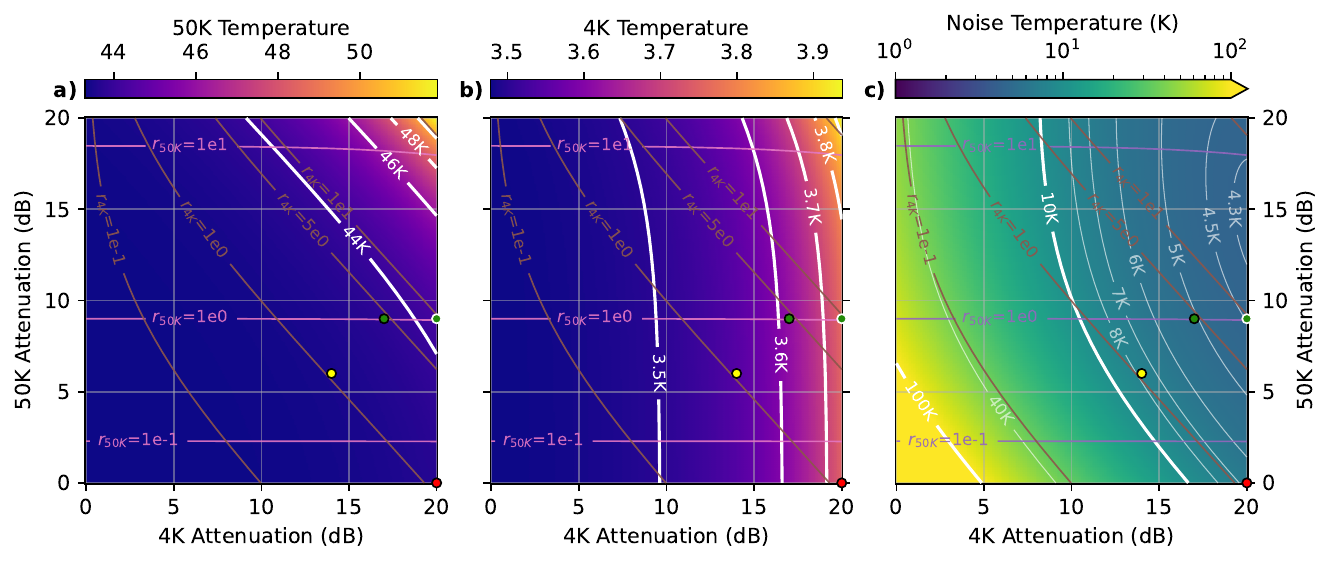}
 	\caption{
        Analysis of flux-line performance for varying noise ratios in the flux line configuration space for our BlueFors LD-400 fridge, with the Still, CP, and MXC attenuators fixed at $0\,\text{dB}$.    
        Panels show plate temperatures for the 50K (a) and 4K (b) stages, and total output effective noise temperature (c).
        Contours identify the corresponding noise ratios, with $r_{{\rm JN},\,50K}$ (purple lines) and $r_{{\rm JN},\,4K}$ (green lines) are overlayed, respectively.
        Selected configurations are shown as circular markers for our systematic, direct optimisation from Fig.~\ref{fig:DCLines} (yellow circle) and existing best practice~\cite{Krinner2019} (red circle).
        Two new points motivated by inverted amplifier-chain design principles are also shown, corresponding to noise ratio values for $r_{{\rm JN},\,50K}$ and $r_{{\rm JN},\,4K}$, respectively, of approximately (1,5) and (1,9) (green points):
        Noise is minimised in the 4K-dominant regime, with $r_{{\rm JN},\,4K}>r_{{\rm JN},\,50K}\approx 1$, but is also associated with increased cooling consumption and operating temperatures.
        Some configurations (black-bordered points) are used in performance comparisons in \ref{fig:performance_comparison}.
    }
    \label{fig:new-heuristic-flux}
\end{figure}

As discussed in Section~\ref{subsec:dc}, it is impractical to install any substantive amount of attenuation on the CP or MXC stages for flux lines, and undesirable to do so on the Still stage (since dramatic variations in applied power at the Still would feedback into dramatic variations in fridge cooling efficiencies).
Since flux lines therefore only include attenuators installed on two stages, it is fairly straightforward to explore the entire configuration space ``brute force'' (Fig.~\ref{fig:DCLines}), which limits the need for heuristics and guiding principles.
It is nevertheless instructive to briefly outline how the principles and analogy discussed above can inform flux-line design, and we illustrate this data in Fig.~\ref{fig:new-heuristic-flux}.

The amplifier chain analogy accurately represents the flux-line scenario for several reasons (which we will see do not apply to drive lines): 
1) Flux-line control signals operate at comparatively low frequencies;
2) The Johnson-Nyquist cut-off frequency for the lowest temperature stage where attenuation is installed is substantially higher than flux-control frequencies;
3) The temperature changes induced by flux-line control signals on the 50K and 4K stages are relatively modest, due to their large cooling powers.
Consequently, the noise defined by Eq.~\ref{eqn:attenuation_output_noise} is relatively independent of temperature and frequency, and determined primarily by the chosen attenuation values.

An na\"ive ``equal balance'' design heuristic, based on the target operating temperatures from Table~\ref{tab:cooling_budget}, would recommend approximately $8\,\text{dB}\approx 300\,\text{K}/46\,\text{K}$ of attenuation on the 50K plate and around $11\,\text{dB}\approx 46\,\text{K}/3.86\,\text{K}$ on the 4K plate, totalling roughly 18--20$\,\text{dB}$, consistent with the maximum attenuation recommended in~\cite{Krinner2019}.
(When simulated operating temperatures and the $1-1/A$ term are used, this becomes around $9\,\text{dB}$ and $11\,\text{dB}$ on the 50K and 4K plates, respectively.)
At this point, the contributions of the 4K, 50K and RT stages are all approximately equal, giving a total noise around three (3) times larger than the noise floor set by the 4K stage (with the 4K plate $\sim3.5\,\text{K}$ and $A_\text{4K}=11\,\text{dB}$).

A heuristic based on an inverted amplifier chain, however, would advise operating in the 4K-dominant regime, with ratios, say, of $r_{{\rm JN},4K} \approx 4.5 > r_{{\rm JN},50K} \approx 1$, corresponding to an attenuation configuration of $A_\text{50K}=9\,\text{dB}$ and $A_\text{4K}=17\,\text{dB}$ indicated by the green point in Figure~\ref{fig:new-heuristic-flux} (9-17-0-0-0).
By moving some attenuation from the 4K plate to the 50K plate, and then adding additional attenuation on the 50K plate, this delivers a substantial improvements over the 0-20-0-0-0 solution recommended in~\cite{Krinner2019}, with around $25\%$ reduction in total noise, and over $40\%$ reduction in heat load on the 4K plate.
Importantly, this required removing the constraint of a fixed total attenuation, which represents a key difference from the standard ``equal balance'' design heuristic.
Both noise reduction and increased heat load are even larger compared with the 6-14-0-0-0 target configuration identified through empirical modelling in the first part of this paper, again a result of removing the constraint on total attenuation.
Returning the 4K attenuation to $A_\text{4K}=20\,\text{dB}$ (which also returns its associated heat load), still with $A_\text{50K}=9\,\text{dB}$ (9-20-0-0-0), almost doubles the 4K noise ratio to $r_{{\rm JN},4K} \approx 9.3 > r_{{\rm JN},50K} \approx 1$, but only gives a further $\sim 6\%$ improvement in overall noise temperature relative to the 0-20-0-0-0 solution.

Adding flux-line attenuation on the 50K plate obviously adds substantial active flux load to that temperature stage.
For our line configuration, this is well within the fridge's cooling capacity, increasing the temperature of the 50K stage by less than a kelvin.
The most significant drawback to the increased total attenuation recommended by this approach, however, is currently likely to be the significantly increased flux voltage required at the flux-line inputs at the top of the fridge, a trade-off that depends on factors currently external to our model.
Thus, the optimal approach may ultimately be determined by a compromise between what maximum pulse voltages can be produced by the fast flux-pulsing electronics outside the fridge, and what mutual inductance (that is, the transmon junction loop size) can be tolerated without introducing too much magnetic flux noise from other sources (as opposed to the Johnson-Nyquist noise considered here).
Constraints such as these also need to be modelled and monitored for cryogenic quantum processors where capacity is a constrained resource.

\subsection{Guiding Principles for Attenuator Design: Drive Lines}

Unfortunately, realising the MXC-dominant regime can be quite difficult in practice for drive lines which reach deep into the cryogenic regime ($hf \gg k_\text{B}T$).
A face-value application of amplifier-chain guiding principles does not automatically account for the significant heat-load-dependent fridge temperature response which arises for the attenuations involved.
While the heat loads created to reach the 4K-dominant regime for flux lines produce only modest variations in 4K and 50K plate temperatures, this is not the case to reach the MXC-dominant regime for the drive lines, especially for the MXC stage.
As discussed above, cut-off frequencies for $21\,\text{mK}$ MXC and $130\,\text{mK}$ CP stages are around $440\,\text{MHz}$ and $2.7\,\text{GHz}$, respectively, making attenuator noise exponentially sensitive to temperature changes at typical circuit QED operating frequencies of 4--$8\,\text{GHz}$, especially for the MXC.
Since the signal power needed at the device is fixed, the MXC heat load increases in proportion to the MXC attenuation; once the active drive contribution becomes significant relative to other heat load contributions at the MXC, increasing the MXC attenuation directly influences the MXC operating temperature.
In Eq.~\ref{eqn:attenuation_output_noise}, the influence of stage temperatures on output noise enters through the Boltzmann-distributed Johnson-Nyquist noise terms, $S_{\rm i}(T_i, f)$ (Eq.~\ref{eqn:psd_JN}).
In the amplifier-chain analogy, this corresponds to the first amplifier's input noise increasing proportionally as its gain is turned up.
Consequently, care needs to be taken to correctly identify the guiding principles that apply to drive-line attenuation design, and this depends crucially on the full, detailed fridge response.

\subsubsection{A Modified ``Equal-Balance'' Principle for Drive Lines with Fixed Total Attenuation}

\begin{figure}[!htp]
	\centering
 	\includegraphics[width=\linewidth]{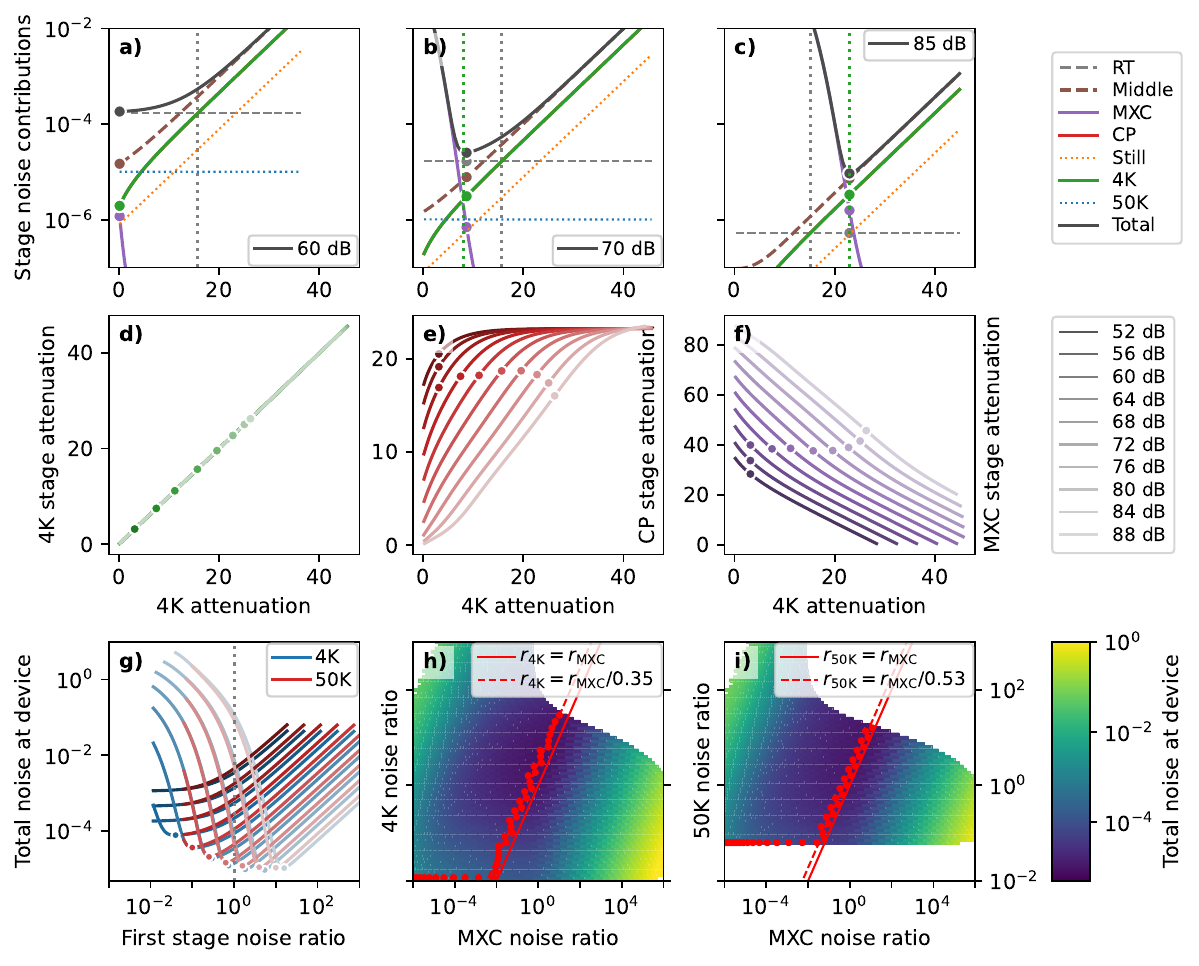}
 	\caption{
        Characterisation of drive-line performance for three-stage attenuator configurations (4K-CP-MXC) for varying noise ratios ($r_{{\rm JN},\,{\rm MXC}}$ and $r_{{\rm JN},\,i \neq {\rm MXC}}$) for fixed total attenuations.
        Attenuations on the 4K (d) and CP (e) stages are chosen to contribute equal noise at the device, and any leftover attenuation is allocated to the MXC (f).
        Each curve (``vertical'' slices in (h, i)) spans a range of 4K noise ratios between $r_{{\rm JN},4K}=10^{-3}$ and $r_{{\rm JN},4K}=10^{3}$ (filtering out data where the iterative model does not converge to a valid solution) for a particular fixed total attenuation 50 and $90\,\text{dB}$.
        (a--c) Total device noise and contribution breakdown---room temperature, cryogenic stages, and the sum of intermediate stages (``middle'')---plotted against 4K attenuation for three distinct total attenuation regimes: (a) $60\,\text{dB}$, the noise minimum being at the limit of where the model iteration converges; (b) $70\,\text{dB}$ and (c) $85\,\text{dB}$, the noise minimum being located before and after the full ``equal-balance'' point, respectively.
        Filled circles indicate where minimum total noise occurs---the locations consistent with a modified ``equal-balance'' principle---and vertical dotted lines indicate where middle contributions equal the RT contribution (not, in general, aligned with minimum noise).
        (d--i) Performance for a full range of total attenuations.
        (d--f) Three-stage configurations for 4K, CP and MXC attenuations, respectively, comparing a wider selection of total attenuation budgets.
        (g) Comparison of total device noise for three-stage (4K-CP-MXC) and four-stage (50K-4K-CP-MXC) configurations, illustrating they achieve comparable minimum noise at comparable noise ratios.
        (h,i) Full 2D landscapes plotting device noise for three- and four-stage configurations against first- and final-stage noise ratios.
        The solid red curve, \emph{not a fit}, shows noise is near optimal when all stage noise contributions are matched, with the dashed line scaling the MXC ratio by a multiplicative factor calculated as a geometric average over points in the appropriate region.
    }
    \label{fig:drive_line_heuristic}
\end{figure}

To understand the impact of attenuator configuration on drive-line performance for different regimes in relation to the relative noise contributions from different temperature stages, we first consider different fixed total attenuations and sweep the noise ratios according to Eq.~\ref{eqn:attenuation_output_noise_ratios}.
To do this, we use our full holistic fridge model to simulate fridge performance for a given attenuator configuration, and use Equations~\ref{eqn:attenuation_output_noise} and~\ref{eqn:psd_JN} for the simulated stage temperatures to iteratively adjust the attenuations required to achieve a specified noise ratio for the intermediate stages, with unallocated attenuation at each iteration applied to the final stage.
This adaptive process thus identifies the temperature-dependent attenuator configuration which gives rise to a specific distribution of noise contributions across the fridge stages.
This adaptive configuration design is enabled crucially by our detailed characterisation of the full fridge temperature response (see App.~\ref{app:temp_response}).

The first row, Figs~\ref{fig:drive_line_heuristic}~(a--c), shows the noise contributions for each stage (including RT) as a function of 4K attenuation (the first stage with attenuation), along with the total noise at the device, and the aggregate noise from all intermediate stages (here, 4K and CP), for total attenuations of $60\,\text{dB}$, $70\,\text{dB}$ and $85\,\text{dB}$.
The second row, Figs~\ref{fig:drive_line_heuristic}~(d--f), shows the attenuation configurations for each temperature stage (here, constrained to the 4K, CP and MXC plates, with no attenuation on either 50K or Still plates) for a range of different total attenuations between $52\,\text{dB}$ and $88\,\text{dB}$.
As noted previously, Equation~\ref{eqn:noise-ratio-adjacent} shows that the noise ratio between any two adjacent plates is determined primarily by the attenuation on the lower plate, and depends only weakly on the upper attenuation.
This means that changing the noise ratios for all intermediate stages, $r_{{\rm JN},i\neq\text{MXC}}$, is achieved primarily by changing the attenuation on the first plate, with attenuation remaining roughly constant on the following intermediate plates.
Because total attenuation is fixed, changes in the first attenuator is offset by balancing changes on the final attenuator.

\begin{namedprinciple}[\emph{Modified ``Equal-Balance'' Principle for Fixed Total Attenuations:}]
\emph{For drive lines with fixed total attenuation and reaching into the deep cryogenic regime ($k_B T \ll hf$ for the lowest temperature stage with attenuation), the attenuator configuration will achieve near-optimal total device noise when the noise contribution of the lowest temperature stage approximately matches the noise contributions from the other intermediate stages.  
At low frequencies or high temperatures, the modified ``equal-balance'' point will achieve a total device noise very close to the minimum possible device noise (although the location of that true minimum may not be all that close to the ``equal-balance'' point, due to saturation of the noise floor).}    
\end{namedprinciple}

The results in Figs~\ref{fig:drive_line_heuristic}~(a--c) illustrate that the total minimum noise does not occur where the MXC contribution is the sole dominant contribution, as would be expected from a na\"ive application of the reverse amplifier chain principle, but rather where the MXC noise ratio balances the intermediate stage ratios, defining a modified or restricted ``equal-balance'' principle.  At different fixed total attenuations, the minimum total noise, marked by filled circles, occurs very close to the configuration where all stage contribution ratios are equal, $r_{{\rm JN},MXC} = r_{{\rm JN},i\neq\text{MXC}}$, roughly indicated by vertical green dotted lines.
(There is a small, systematic offset consistent with the MXC noise contribution exhibiting a steep scaling vs 4K attenuation due to the high MXC attenuations needed, by comparison with the intermediate stage noise contribution scalings.)
Furthermore, the results show that this restricted principle holds true for balanced noise ratios both smaller and larger than one (1) (corresponding to less and more fixed total attenuation, respectively).
Moving either way from this point shifts attenuation to or from the MXC stage (Figs~\ref{fig:drive_line_heuristic}~(d--f)):
shifting attenuation from 4K to MXC reduces the intermediate noise contributions $r_{{\rm JN},i\neq\text{MXC}}$, but increases the MXC temperature, and hence the MXC noise $r_{{\rm JN},MXC}$; while shifting more attenuation to the higher stages so that $r_{{\rm JN},i\neq\text{MXC}} > r_{{\rm JN},MXC}$, reduces the MXC temperature and corresponding noise, but increases the higher stage contributions.
Because the RT contribution depends only on total attenuation, not the configuration, it is flat and does not affect the location of the minimum.

Results from our earlier drive-line analysis in Figure~\ref{fig:ACLines} are also consistent with a restricted ``equal-balance'' heuristic, with the total noise minima occurring when the MXC contribution was equal to the other variable stage contributions.
Specifically under the fixed total attenuation of $60\,\text{dB}$, the minimum noise occurs at $A_\text{4K}=0.3\,\text{dB}$, $A_\text{CP}=12\,\text{dB}$ and $A_\text{MXC}=47.7\,\text{dB}$, which corresponds to contribution ratios of $r_{{\rm JN},MXC} \approx r_{{\rm JN},i\neq\text{MXC}} \approx 0.1$.
Figures~\ref{fig:drive_line_heuristic}~(e,f) show that, at low values of total attenuation, the CP and MXC ``equal-balance'' attenuations diverge sharply from the general trend, aligning with a point where the 4K attenuator reaches a constant lower bound connected with intrinsic cable attenuation.

Figure~\ref{fig:drive_line_heuristic}~(g) shows the full noise seen by the device for the same total attenuations plotted in Figs~\ref{fig:drive_line_heuristic}~(d--f), and compares the results for three- and four-stage configurations (4K-CP-MXC vs 50K-4K-CP-MXC).
Because of the lower temperature difference between 50K and room temperature, our model cannot find solution configurations for noise ratios as low as for the three-stage case.
More importantly, however, whenever both three- and four-stage converge to minimum-noise solutions, they achieve comparable total device noise.
Three-stage configurations achieve slightly lower noise minimums at the cost of increased 4K heat loads, due to the increased attenuation at 4K (the CP and MXC attenuations being comparable for both types).
Figure~\ref{fig:drive_line_heuristic}~(h,i) show full, higher resolution landscapes of total noise versus first and final stage noise ratios.
The linear distribution of the minimum data points in each curve (shown by red circles) in close proximity to the solid $r_{{\rm JN},MXC} = r_{{\rm JN},i\neq\text{MXC}}$ line (or more specifically that $r_{{\rm JN},MXC} \sim 0.35\,r_{{\rm JN},4K}$ and $r_{{\rm JN},MXC} \sim 0.53\,r_{{\rm JN},50K}$ for three-stage and four-stage configurations, respectively, independent of the total fixed attenuation), verify that a modified ``equal-balance'' heuristic applies, for both three- and four-stage configurations, across a large range of total attenuations.

For a fixed total attenuation, these results set a clear universal target for how to distribute attenuation across the stages, with the optimal value for the contribution ratios depending on the chosen total attenuation.
They also show that the minimum achievable noise occurs when the now-equal stage contributions dominate over the RT contribution, $r_{{\rm JN},MXC} = r_{{\rm JN},i\neq\text{MXC}}>1$:
for $r \gtrsim 1$, increasing total attenuation decreases the RT contribution proportionately, while affecting the stage contributions only minimally, being realised almost entirely by increased attenuation at the first thermalised stage.
While this is consistent with inverted amplifier chain design principles, the results in Fig.~\ref{fig:drive_line_heuristic}~(g) nevertheless show only fractional improvement beyond $r\sim1$ for the operational parameters we studied, since only the RT noise contribution is further suppressed (one out of the four or five stages) while the plates either hold the same or increased temperature due to excessive heat loads.

These results extend the analysis in Section~\ref{subsec:ac}, and our insights in relation to standard practice (e.g.,~\cite{Krinner2019}), in three key ways:
\begin{itemize}
\item Firstly, our results ultimately support an intuition that heat loads and noise suppression for drive lines are balanced effectively when noise contributions are matched between fridge stages, but also clearly show that implementing this in practice depends critically on the full fridge response---only possible through holistic modelling of the interplay between fridge temperatures, heat loads, and attenuations.  None of these factors are usually considered in the standard, simplistic way that ``equal balance'' ideas are applied.
\item Secondly, our analysis also shows that the fairly ubiquitous practice of choosing a fixed total attenuation early in the design process---which practice we adopted in the first part of this paper, to constrain and simplify an otherwise complex optimisation problem with many degrees of freedom, and facilitate direct performance comparison against standard community practice---should potentially be relaxed to achieve noise improvements of potentially more than an order of magnitude.
\item Thirdly, matching the MXC noise contribution to the other stages does typically require very substantial attenuations and associated cooling budgets, compared with conventional aims to keep MXC heat loads rather low, but tolerating increased cooling budgets at the MXC stage may nevertheless be a worthwhile trade-off, given the potential improvements in noise performance, depending on their impact on other factors not yet included in our fridge model.
\end{itemize}

\subsubsection{A Full ``Equal-Balance'' Guiding Principle for Drive Lines without Constraints on Total Attenuation}

\begin{figure}[!htp]
	\centering
 	\includegraphics[width=\linewidth]{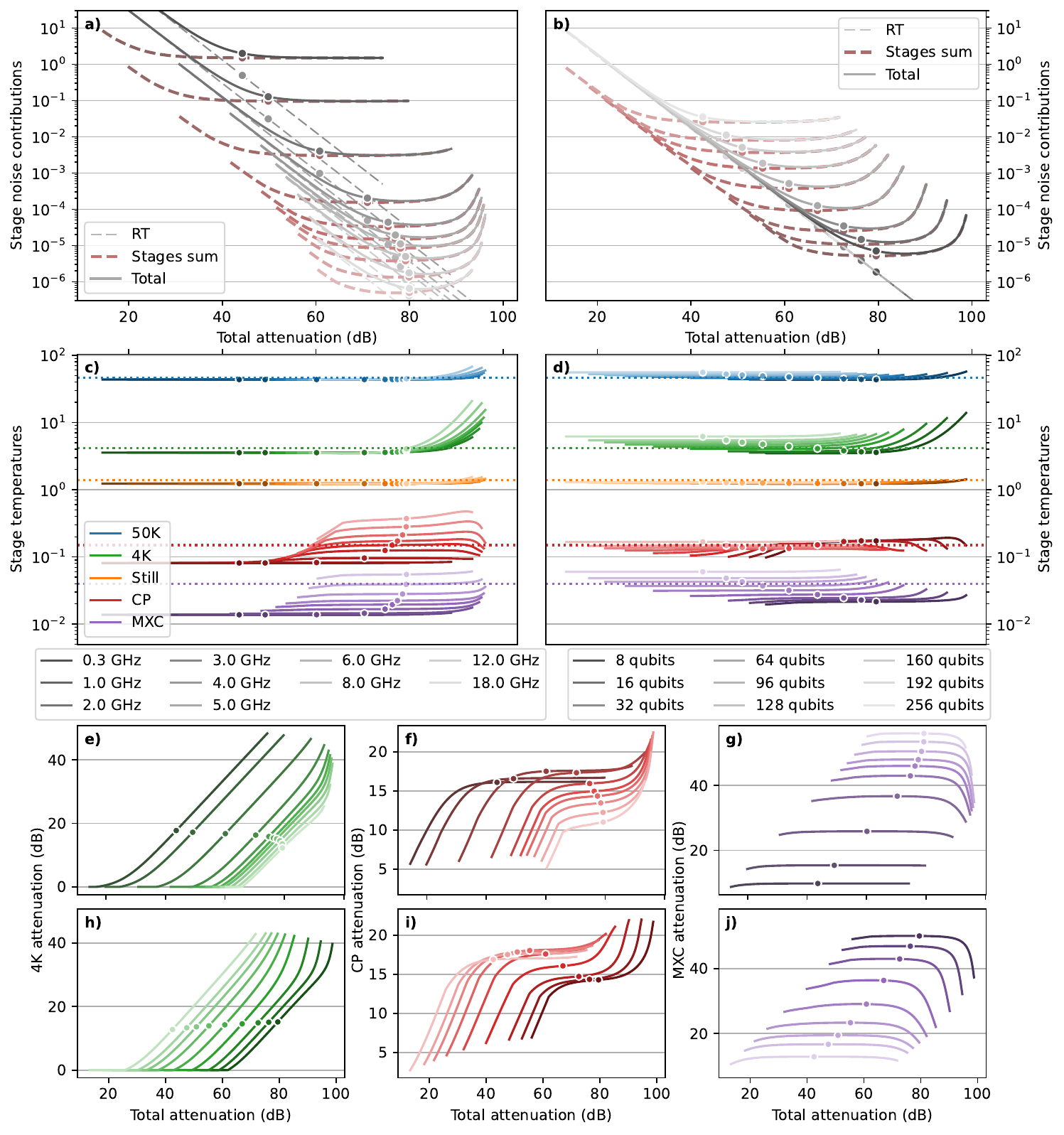}
 	\caption{
        Characterisation of drive-line performance for three-stage attenuator configurations (4K-CP-MXC) satisfying the modified ``equal-balance'' principle under different operating conditions (for operating frequency and qubit control capacity):
        Attenuations on all cryogenic stages are chosen to contribute equal noise at the device, $r_{{\rm JN},\,{\rm MXC}}=r_{{\rm JN},\,i \neq {\rm MXC}}$, with specified noise ratios relative to room temperature (not necessarily 1), with total attenuation dependent on and determined by the noise ratio.
        Plots show drive-line and fridge performance for 0.3--$18\,\text{GHz}$ in frequency and from 8--256 qubits, with data plotted for a range of noise ratios between $r_{{\rm JN}}=10^{-3}$ and $r_{{\rm JN}}=10^{3}$ (filtered where the iterative search does not converge).
        (a,b) Noise at the device, along with contributions from RT and an aggregate from all cryogenic stages, against total attenuation, for different frequencies (a) and number of qubits (b).
        (c,d) Plate temperatures, with dotted horizontal lines indicating target operating temperatures, and dashed lines marking selected operational limits ($4.2\,\text{K}$ for the 4K stage, $1.3\,\text{K}$ for the Still, and $40\,\text{mK}$ for the MXC, where thermal excitation $p\approx10^{-3}$).
        Attenuation distributions required to satisfy the full ``equal-balance'' principle are shown for the 4K, CP and MXC plates, respectively for (e--g) different operating frequencies and (h--j) different numbers of qubits.
        The full ``equal-balance'' points ($r_{{\rm JN},MXC} = r_{{\rm JN},i\neq\text{MXC}}=1$), where all noise contributions are equal (including from RT) are indicated by shaded circles.
    }
    \label{fig:operating_parameters_sweep}
\end{figure}

The previous section shows that, if drive-line configurations are optimised subject to a fixed constraint on total attenuation, optimal noise is achieved when all cryogenic noise contributions are balanced with each other, but not necessarily with the room-temperature contribution.
Comparing data across different choices of total attenuation, however, suggests that choosing a fixed total attenuation may significantly constrain achievable performance.
To explore this more comprehensively, we therefore relinquish the standard constraints on total attenuation, and instead restrict attention to only drive-line configurations that match the modified ``equal-balance'' principle, $r_{{\rm JN},MXC} = r_{{\rm JN},i\neq\text{MXC}}$ $\forall\,i$, as defined above.

Applying the modified ``equal-balance'' principle identifies a unique, full drive-line attenuator configuration for each choice of a single noise ratio value, subject to the fridge temperature response and which stages are chosen to accommodate attenuators.
Constraining the configurations to these optimal design points thus greatly reduces the free parameters required to explore some search space of interest.
For example, the temperatures, attenuations and heat loads which determine the fridge response, \emph{are} significantly affected by higher-level fridge operating parameters (e.g., operating frequency, number of qubits, cable materials, etc).
This method for identifying near-optimal attenuator configurations therefore enables a more comprehensive, bigger picture exploration of fridge performance, and we explore this over a wide range of potential operating regimes in relation to frequency and qubit capacity (as a proxy providing context for choosing a number of control lines, with 1 drive and 1 flux line per qubit, and 1 readout input and output line per 8 qubits).
Numerical results from our full, holistic fridge model for 4K-CP-MXC three-stage configurations, under variations in both of these parameters is shown in Figure~\ref{fig:operating_parameters_sweep}.

The first two panels, Figs~\ref{fig:operating_parameters_sweep}~(a) and~\ref{fig:operating_parameters_sweep}~(b), show the total noise seen by the device, as a function of total installed attenuation (not including intrinsic cable attenuation), for different operating frequencies and number of qubits, respectively.  
Each solid line shows the performance achieved for a wide range of noise ratios spanning between $10^{-3}$ and $10^{3}$ for $r_{{\rm JN},MXC} = r_{{\rm JN},i\neq\text{MXC}}$\footnote{A small number of curves have been truncated to display only regions where the iterative fridge model was able to converge successfully to a solution, but this was a very limited number and does not impact any conclusions, so we did not try to mark this in any way, to avoid crowding the plots.}, defining the modified ``equal-balance'' heuristic, with the full ``equal-balance'' heuristic point ($r_{{\rm JN},MXC} = r_{{\rm JN},i\neq\text{MXC}}=1$) marked with circular markers.
Panels (a) and (b) also explicitly show the RT noise contribution (grey dashed line) and the aggregate noise contribution from all cryogenic stages (brown dashed line), clearly illustrating how the total device noise is dominated by the RT contribution for noise ratios less than one (1), and by the cryogenic stages for noise ratios greater than one (1).
The next two panels, Figs~\ref{fig:operating_parameters_sweep}(c) and~\ref{fig:operating_parameters_sweep}(d), show the predicted temperature response for each stage.
The dotted horizontal lines mark the target operating temperatures outlined in Table~\ref{tab:cooling_budget}, and the dashed horizontal lines indicate selected operational limits.
The panels (e)--(g) (third row) and (h)--(j) (fourth row) show the attenuations installed on the 4K, CP and MXC plates, for different frequencies and qubit control capacities, respectively.

The first key observation to highlight is that the total attenuations required to reach the full ``equal-balance'' points in practical operational scenarios (e.g., 4--8 GHz and 8--32 qubits) are quite substantial: 70--$80\,\text{dB}$, of which at least 40--$50\,\text{dB}$ is typically installed at the MXC.
This gives rise to quite significant heat loads on the MXC, as well as the CP stage.
Interestingly, while the MXC attenuations required vary strongly with operating parameters, they are remarkably constant for rather widely varying noise ratios, meaning that decreasing the noise ratio is typically not a particularly effective way to reduce the load on the most sensitive MXC plate.
Perhaps surprisingly, however, despite those high attenuation values, they do not immediately translate into prohibitively large fridge temperature responses, with the predicted MXC temperatures appearing to stay below around $25\,\text{mK}$ (or at most $30\,\text{mK}$) for a wide range of operating parameters (see Figs~\ref{fig:operating_parameters_sweep}~(c) and (d)).
These temperatures do not translate into a significant increase in MXC Johnson-Nyquist noise, nor do they give rise to large equilibrium thermal excitations at device frequencies.
This suggests that our fridge can actually tolerate a substantially higher active drive load than our target operating temperatures\footnote{which are consistent with standard operating temperatures identified in other works, e.g.,~\cite{Krinner2019,Raicu2025}} would suggest.
It is important to remember that noise photons and thermal excitations, etc., are not necessarily the ultimate limiting constraints on experimental cryogenic operating conditions.
There may be numerous other practical considerations, not already included in our holistic model, which could set a hard operating limit before such high MXC attenuations are reached, e.g., regarding available control signal powers, pulse speeds and pulse durations.  
Nevertheless, these results suggest that it will be important to investigate these effects, especially in contexts where achievable quantum processor capacity and complexity are critical considerations.

The next key observation is that the noise ratios at which the total device noise is minimised are all substantially larger than the true ``equal-balance'' point, typically around $r\sim10$--50.
Strictly speaking, therefore, this demonstrates that the full ``equal-balance'' heuristic does not technically identify the minimum-noise configuration, which requires significantly more (10--$20\,\text{dB}$) extra total attenuation.
However, while the optimum technically involves significantly more total attenuation than the ``equal-balance'' configuration, in practice it achieves only very minimal improvement in terms of noise.
On the other hand, up to the ``equal-balance'' point in total attenuation, increasing the attenuation delivers relatively strong reduction in noise compared with the additional heat load.
We therefore define the following design heuristic for drive lines which are not constrained in terms of total attenuation:

\begin{namedprinciple}[\emph{Full ``Equal-Balance'' Principle for Fixed Noise Ratios and Variable Total Attenuation:}]
\emph{For drive lines with no restrictions on total attenuation and reaching into the deep cryogenic regime ($k_B T \ll hf$ for the lowest temperature stage with attenuation), full ``equal-balance'' attenuator configurations, with $r_{{\rm JN},MXC} = r_{{\rm JN},i\neq\text{MXC}}=1$ across all stages, will achieve near-optimal total device noise.}
\end{namedprinciple}

From the temperature modelling plotted in Figs~\ref{fig:operating_parameters_sweep}~(c) and (d), we see that increasing the operating frequency and increasing the qubit control capacity both lead to increased MXC heat loads and temperatures.
This behaviour occurs for quite different reasons, however, which can be understood by looking at the total device noise, which increases with qubit control capacity, but \emph{decreases} significantly with increasing operating frequency.
In the case of increasing qubit control capacity, the increased noise is \emph{caused} by the increased operating temperatures across all fridge stages.
Part of this increase arises because the inclusion of more drive lines magnifies the response to drive line attenuation configurations, but there is also a large, fairly uniform background increase from passive loads and active flux loads, which do not depend on the drive line attenuation configuration.
The total fridge heat burden, including through signal dissipation in the attenuators, scales linearly with the number of lines.
As the MXC operating temperature increases, because of the larger heat burdens on all stages, this exponentially reduces the amount of MXC attenuation required to realise the (higher-noise) ``equal-balance'' noise condition; that is, the need for attenuation reduces only because there is reduced underlying room to reduce the device noise.

In the case of increasing operating frequency, however, the roles of cause and effect are almost directly reversed.
As operating frequencies are increased, which has no intrinsic effect on operating temperatures, the performance-limiting noise floor from the MXC attenuator drops rapidly from moving deeper into the cryogenic regime of the Johnson-Nyquist noise spectrum ($hf\gg k_\text{B}T$).
This, however, dramatically increases the attenuations required, particularly on the MXC, to reach the full ``equal-balance'' operating point.
As discussed above, in the end, the ``equal-balance'' point here is achieved, to a significant degree, because of the increased MXC temperatures that result from increased attenuation at the MXC stage.
It is worth noting, however, that quantum components operating at higher frequencies are generally better able to tolerate higher operating temperatures without suffering too badly from the impacts of thermally induced excitations.
Finally, we note that the total attenuations required to reach optimal ``equal-balance'' operation become more moderate as qubit control capacity is increased, or operating frequency is decreased.
While this is also accompanied by generally poorer noise performance, if it is necessary to operate in these regimes for a specific context, our results suggest it is likely to be less challenging to reach and operate using optimal ``equal-balance'' attenuation configurations.

Our analysis (some results not shown) also reveals how the number of attenuators used in the configuration affects performance,
specifically comparing configurations with 4 attenuators (50K-4K-CP-MXC) and 3 attenuators (4K-CP-MXC).
In general, using more attenuators does not appear to substantially impact the achievable noise performance (see, e.g., results in Fig.~\ref{fig:drive_line_heuristic}~(g)), but it can affect the robustness of the noise performance with respect to total attenuation.
This broadly agrees with the analysis in Section~\ref{subsec:ac}, where noise performance was shown to be largely invariant when introducing a 50K attenuator.
At large total attenuation, however, the 3 stage configuration breaks down faster than the 4 stage configuration; the first stage of the four-stage configuration has a smaller temperature response and larger cooling budget to absorb the increased heat load due to signal dissipation.

As noted above, for some more extreme operating parameters, our model failed to converge to attenuations which validly realised a modified ``equal-balance'' configuration.
In general, these failures occurred where Still temperatures exceeded the range captured by our current temperature response modelling, or increased past reasonable operating limits (e.g., dictated by still line pumping pressures).
Despite not having any attenuators installed at the Still plate, some active drive line dissipation still arises from the intrinsic attenuation in the cable connecting the 4K and Still plates.
This is an important factor in determining whether it is viable to push the fridge configuration limits far enough to achieve optimal noise performance, since the Still temperature also has a crucial impact on the available cooling power and temperature at the MXC, where the device is located.
In the course of our analysis, one way we were able to extend the range of parameters for which our model converged to a valid solution (not shown), was to configure the drive lines with a section of superconducting cable between the 4K and Still plates.
Depending on the specific parameter regime a given quantum processor is operating in, it may be that performance or qubit capacity can be extended through strategic use of small SC cable segments between the 4K and Still plates.

Again, these results significantly extend the analysis in Section~\ref{subsec:ac}, and our understanding of optimal attenuation design for drive lines compared with standard practice and widely accepted rules of thumb.  For example:
\begin{itemize}
\item The way that ``equal-balance'' ideas are typically employed in cryogenic cable attenuation design is that: a) relatively conservative target operating temperatures are chosen for each temperature stage; b) a fixed total line attenuation is chosen; c) these temperatures are used to directly calculate the ``equal-balance'' reference attenuation values according to $A_i = S_\text{JN}(T_{i{-}1}, f)/S_\text{JN}(T_{i}, f)$ for stage $i$, which does not include potentially significant variations from the $1-1/A$ term, and which is often (usually) further simplified using the classic Johnson-Nyquist noise formula, $A_i \approx T_{i{-}1}/T_{i}$; and d) the reference values are used to assign attenuations to cryogenic stages from top to bottom, with the final stage being allocated whatever leftover attenuation is remaining from the chosen attenuation budget.   By contrast, focussing on the circular markers which locate the full ``equal-balance'' points in Figs~\ref{fig:operating_parameters_sweep}~(e--j), we can see that the attenuation values required to realise the full ``equal-balance'' configuration vary significantly with the fridge operating parameters.
\item The iterative modelling required to identify attenuator configurations that deliver the target noise ratios, while simultaneously keeping track of the predicted fridge response, illustrates that ``equal-balance'' principles cannot be applied successfully a) without incorporating detailed modelling of the fridge response, b) without removing the constraints on total line attenuation, c) without including all relevant noise corrections (e.g., the $1-1/A$ terms and the full frequency-dependent Johnson-Nyquist noise response, as per Eq.~\ref{eqn:psd_JN}).
\end{itemize}

\section{Conclusion and Outlook}\label{sec:conclusion}

\begin{figure}[!htp]
	\centering
 	\includegraphics[width=\linewidth]{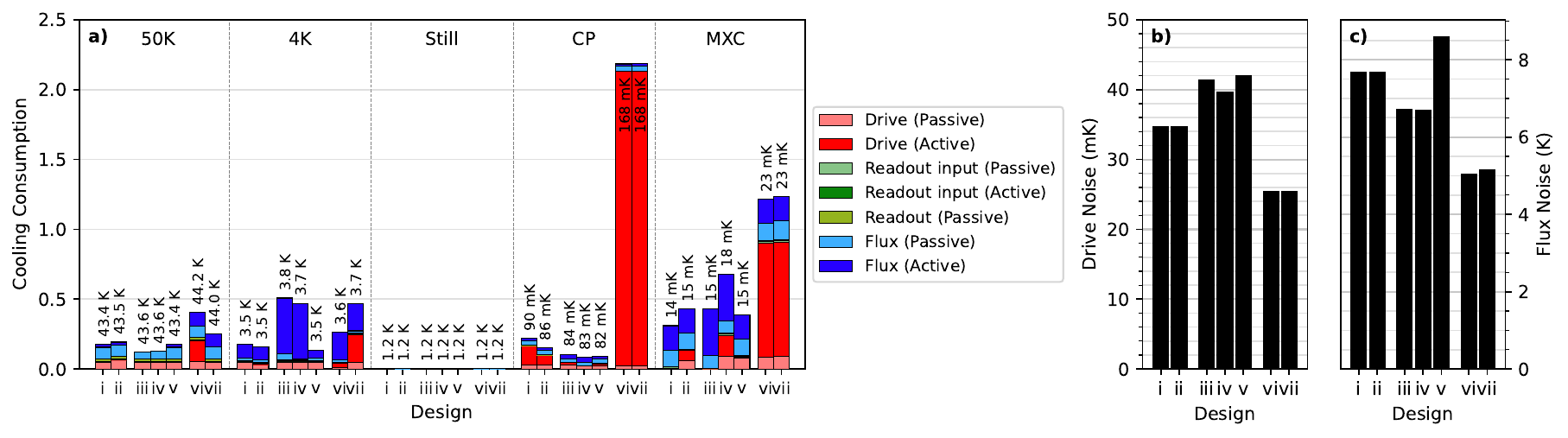}
 	\caption{
        Overall performance of full fridge configurations to compare our optimised designs against existing best-practice designs heuristics.
        Data shown for the LD-400 fridge for the designs indexed in Table~\ref{tab:compared_designs}, including (i, ii) designs from our systematic direct optimisation in Section~\ref{sec:optimisation}, (iii) previous best practice solutions identified in~\cite{Krinner2019}, (iv) a simplified solution using the minimum number of ``active'' stages, (v) the standard, simplistic ``equal-balance'' heuristic used with fixed total attenuation, and (vi, vii) new three-stage and four-stage attenuation configurations satisfying a full, equal-balance heuristic based on the rigorously supported general design principles identified in Section~\ref{sec:new_heuristic}.
        (a) Full fridge heat loads (referenced to target maximum cooling powers specified in Table~\ref{tab:cooling_budget}) and associated plate temperatures estimated from our full fridge model (see App.~\ref{app:temp_response}).
        (b, c) Noise performance for dedicated drive lines (b) and flux lines (c) as effective output noise temperature (noise performance for switchable lines used as drive and flux lines not shown here).
        Our first systematic, direct optimisation has delivered designs with improved performance in device temperatures, qubit capacity (based on cooling consumption), drive line noise, and flexibility.
        Designs based on generalised design principles, which relinquish the constraint on total attenuation, deliver substantial reductions in the output noise, but at the cost of substantially increased cooling consumption and device temperature.
        This may be a worthwhile trade-off, but will likely depend on context, and external factors not currently included in our model. 
    }
    \label{fig:performance_comparison}
\end{figure}

\begin{table}[h]
\caption{
    Outline of design configurations compared in Figure~\ref{fig:performance_comparison}
}\label{tab:compared_designs}
\begin{tabular*}{\textwidth}{@{}p{1.5cm}ccC{0.8cm}C{2.3cm}C{0.8cm}C{2.2cm}@{}}
\toprule
&&Design&Drive Cable&Drive Attenuation&Flux Cable&Flux Attenuation\\
\midrule
\multirow{2}{=}{Systematic Optimisation}&i&Optimised&SS&[0,10,0,20,30]&CuNi&[6,14,0,0,0]\\
&ii&50\% Switchable&CuNi&[6,14,0,0,40]&CuNi&[6,14,0,0,0]\\\hline
\multirow{3}{=}{Existing Heuristics}&iii&Best Practice~\cite{Krinner2019}&SS&[0,20,0,20,20]&SS&[0,20,0,0,0]\\
&iv&2-stage&SS&[0,20,0,0,40]&SS&[0,20,0,0,0]\\
&v&``Equal Balance''&SS&[8,12,0,20,20]&CuNi&[8,12,0,0,0]\\\hline
\multirow{2}{=}{New Heuristic}&vi&4-stage&SS&[17,9,0,14,48]&CuNi&[9,17,0,0,0]\\
&vii&3-stage&SS&[0,16,0,14,48]&CuNi&[9,17,0,0,0]\\
\botrule
\end{tabular*}
\footnotetext{SS = 2.19mm diameter stainless steel, CuNi = 2.19mm diameter copper-nickel coaxial cable}
\end{table}

Through a systematic approach to cryogenic wiring optimisation that uses holistic numerical modelling, we have identified multiple opportunities to substantively improve the performance of our cryogenic wiring stack, compared with blindly applying existing design heuristics.
This includes improvements in device temperature, noise temperature, and qubit capacity, which are critical metrics for the performance of quantum computing systems.

Summarising the results presented in this paper, in Figure \ref{fig:performance_comparison} we examine the performance of different cryogenic wiring configurations as applied to the context our BlueFors LD-400 dilution refrigerator system, comparing established best-practice designs and heuristics against design outcomes optimised using our holistic numerical modelling through both systematic direct optimisation and our new generalised design principles.
We simulate the overall performance of our system for each configurations using the same number and combination of lines, signal powers, duty cycles, and target cooling budgets, as described in Section \ref{sec:optimisation}.
The specific configurations are outlined in Table \ref{tab:compared_designs}: Both our systematically optimised designs (i and ii) and existing heuristics (iii, iv, v) were constrained to satisfy total attenuation budgets of $20\,\text{dB}$ and $60\,\text{dB}$ for flux and drive lines, respectively, for fair direct comparisons;
by contrast, the configurations identified using our new generalised design principles were instead constrained to target noise contributions at each stage satisfying specified noise ratios, with total attenuation left to vary accordingly.

The design resulting from our systematic direct optimisation approach (design i) delivers clear benefits compared to the ``best practice'' solutions chosen in~\cite{Krinner2019} (design iii).
The increase in flux-line heat load on the 50K plate from 5\% to around 10\%, and hence flux noise temperature from $6.7\,\text{K}$ to $7.7\,\text{K}$, were strategically traded off for a four-fold reduction of the flux-line heat load at the 4K plate.
In the context of qubit capacity optimisation, this trade-off is a critical element for maximising the number of qubits that can be operated within a reasonable cooling budget, to avoid a substantial total heat load on the 4K plate being the limiting factor for qubit capacity.
With respect to the drive-line configuration, shifting more attenuation to the MXC stage enabled a 16\% reduction in drive-line noise temperature, with the MXC cooling consumption from the drive lines increasing only from 0.2\% to 1.4\%, and the CP cooling consumption increasing from roughly 5\% to 17\%.
Such an increase is justifiable, since the MXC cooling consumption is still dominated by other contributions (flux-line loads), and while not insignificant for the CP, it still leaves the CP operating well within comfortable cooling budgets.
To mitigate the flux loads on the MXC, stainless steel cables have been replaced by more electrically conductive CuNi cables in the flux lines, where a modest increase in passive load is offset by larger reduction in active load (from 33\% to around 17\%).

Overall, our optimised design leaves the device operating at a 10\% lower device temperature, and delivers improvements in both drive line noise temperature and greater potential line capacity, all critical metrics for quantum computing system performance.
In addition to this, the reduced heat loads at the MXC and 4K plates provide capacity to switch to an overall wiring configuration which allows half of the lines to be installed according to a ``switchable'' design (design ii) which trades a degree of performance for increased flexibility in relation to control signal type (see Section \ref{subsec:transformable}), while still maintaining or exceeding established best-practice performance across most domains.
By comparison, the optimal attenuator configurations (designs vi and vii) far exceed the fixed total attenuation budgets applied to the other designs (by 20--$30\,\text{dB}$ for drive lines and $8\,\text{dB}$ for flux lines), with the resulting heat loads sitting at around 220\% and 120\% of the target cooling power budgets for CP and MXC stages, respectively, largely due to active drive-line loads.
Yet despite this substantial heat burden, this configuration delivers almost 40\% and 25\% improvement in noise temperature for the drive lines and flux lines, respectively, when compared against the best-practice configurations identified in~\cite{Krinner2019}, while the device temperature---around $23\,\text{mK}$---is still not much higher than typical operating temperature ranges for superconducting qubit processors.
It delivers over 25\% in drive-line noise over even design i, which was optimised primarily on drive noise, within the fixed $60\,\text{dB}$ total attenuation budget.

Some of the design optimisations we have identified in this work appear to contradict conventional wisdom.
For example, we identified that the use of $0\,\text{dB}$ attenuators is largely irrelevant for drive lines, and has significant practical limitations for flux lines. 
Furthermore, we have also shown that significant benefits can be achieved through the use of low-loss coaxial cables in flux lines to reduce dominant flux biasing active loads, including quantifying the system-wide benefit of employing superconducting cables for all cables, though this design choice is often avoided for cost reasons.
Finally, our results demonstrate that design choices are not generally universally optimal, and must instead be considered in the context of specific system requirements. 
Our systematic approach identifies different preferred outcomes for different device and system designs and constraints, and given different performance priorities such as optimising for flexibility (e.g., for prototype devices) or optimising directly for qubit capacity.
In each case, we identify solutions that perform better than standard practice heuristics, highlighting the benefits that can be achieved through systematic, holistic optimisation.

We next aimed to develop a basis for understanding our design outcomes in more general terms.
We first developed a generic analytical framework for noise propagation through an attenuator cascade, and show how it can be interpreted using well-established design principles for low-noise amplifier chains. 
We then employed this framework to define a simple parametrisation of our holistic fridge model, which enables an efficient wider numerical exploration of attenuator configurations, that we use to develop more generic design heuristics.
Not only does this provide a more general basis for understanding the improvements identified through our first systematic approach, but it also reveals that driving a cryostat more aggressively relative to its operational limits can potentially deliver significant gains in noise suppression.

Specifically, we first confirm that near-optimal noise performance for lines operating in the deep cryogenic regime ($hf\gg k_{\rm B}T$) can be achieved using a full ``equal-balance'' configuration.
Importantly, however, while this is conceptually somewhat similar to traditional matched thermalisation principles~\cite{Krinner2019}, we show that the ``equal-balance'' configuration varies substantively with the specific operating context, and can only be identified accurately through full, holistic modelling of the fridge's interacting responses to attenuator values, system-wide heat loads, and cryostat plate temperatures, none of which is generally accounted for in standard practice heuristics.
We also show that reaching optimal noise performance often requires substantial attenuations and associated heat loads, especially at the MXC stage, exceeding typical target attenuation and cooling power budgets, often very significantly.

While our current modelling suggests that cryogenic quantum processors and their supporting infrastructure can potentially tolerate higher heat loads than previously assumed, and that there is potential value in relaxing the conventionally strict constraints on total attenuation, an overarching lesson from our analysis is that finding the optimal cost-benefit trade-offs will ultimately depend on many details of system parameters and constraints.
For example, our results illustrate that the expected operating temperatures and noise floors at the ``equal balance'' point are strongly dependent on the operating frequency and qubit capacity.
Beyond that, however, finite bandwidths and voltage output ranges for microwave and flux control hardware, in conjunction with Rabi frequency targets for fast qubit gates, can limit the total installable attenuation.
This balance between control speed and noise suppression also depends on the specific device design---including specific driving mechanisms and the design capacitance or mutual inductance coupling the control lines to the qubits---which determines interaction strengths for both signals and noise.
The relative importance of device temperature to effective line noise temperature can also depend on design specifics, such as the specific decoherence mechanisms and control protocols used:
Devices which rely on direct qubit driving may be more sensitive to control line noise temperature, while the maximum fidelity of qubit reset protocols that depend on thermalisation during idling is limited by equilbrium thermal qubit excitations set by device temperature.
Finally, it may also be necessary to leave headroom in the cooling power to accommodate operational variations over long cooldowns (where cooling powers or temperature responses may degrade or vary over months or years), or to accommodate the need to vary control signal duty cycles and powers during cooldowns.
While many of these are not yet included in the numerical modelling presented here, the results we have obtained already suggest that incorporating more of these details into future modelling could further enhance design outcomes.
Ultimately, achieving well-informed design outcomes that optimally balance performance trade-offs for cryogenically demanding quantum processing contexts, will likely only be possible through holistic numerical modelling such as presented here.

Towards the goal of providing a simplified design process that allows for rapid evaluation of designs under different contexts, we have also developed an open-source web-based GUI tool, Cryowala~\cite{Cryowala2024}.
Cryowala allows systems engineers to input their own system parameters to quickly generate a comprehensive summary of expected performance for their cryogenic wiring configuration.
It already has preliminary features to support the same powerful numerical simulations presented in this work, and can be easily extended to include more complex analysis features.

In the context of the rapid scale-up of quantum processors and the increasing demand for more complex and high-performance cryogenic wiring systems, all avenues with the potential to increase the amount of quantum computing resources available for a given computation are being explored.
We suggest the use of systematic, full-system modelling for design optimisation is a valuable and widely applicable  tool that can provide gains across multiple performance metrics, including the noise burden on the device, as well as the qubit capacity that can be supported by a single fridge, where even modest gains can offer significant value.
We emphasise that the benefits of this approach should extend much further than the improvements in device and noise temperature achieved here for a modest dilution refrigerator system: 
This kind of approach can also help identify optimisations for cryogenic control wiring that can help increase the computational power, and reduce resource consumption, for large-scale quantum computing systems.

\backmatter

\bmhead{List of Abbreviations}
RT: Room Temperature, 50K: 50K stage, 4K: 4K stage, Still: Still Plate, CP: Cold Plate, MXC: Mixing Chamber, HEMT: High Electron Mobility Transistor, SC: Superconducting, CuNi: Copper-Nickel, SS: Stainless Steel, JN: Johnson-Nyquist, GUI: Graphical User Interface.

\bmhead{Acknowledgements}

We would like to thank Jared Cole for useful discussions and encouragement,
and Vera Hansper and Ari Riihim\"aki for discussions and support on collecting and analysing temperature response data from our BlueFors LD-400 dilution refrigerator.
We would also like to acknowledge John Lee and Mitchell Lee for their significant contributions to the development of the Cryowala web-based GUI tool.

\bmhead{Authors' information}

Juan Pablo Dehollain is now affiliated with Emergence Quantum, Sydney Nanoscience Hub, Camperdown, NSW 2050, Australia.

\section*{Declarations}

\bmhead{Availability of data and materials}

The datasets used and/or analysed during the current study are available from the corresponding author on reasonable request.

\bmhead{Competing interests}

The authors declare no competing interests.

\bmhead{Funding}

This work has been funded by the Australian Research Council Future Fellowship (FT170100399) and Australian Research Council Discovery Project (DP210101367) of NKL.
AD and GG acknowledge the support from the Australian Government's Research Training Program (RTP), and Sydney Quantum Academy PhD scholarships.
AD also acknowledges the support from the University of Technology Sydney Research Excellence Scholarship.
JPD acknowledges support from the University of Technology Sydney, Chancellor's Postdoctoral Research Fellowship (UTS, CPRDF).

\bmhead{Author's contributions}

AD developed and carried out all numerical modelling, and performed the systematic analysis and data analysis, under the supervision of and in collaboration with JPD and NKL, and additional contributions from GG.
AD carried out all measurements of the temperature response of the dilution refrigerator, with assistance from GG.
AD and NKL developed, modelled and analysed the analytical framework for attenuator cascades, and wrote the manuscript with contributions from JPD.
The project was led and supervised by NKL. 
All authors read and approved the final manuscript.

\begin{appendices}

\section{Default wiring configurations}\label{app:default-configurations}

\begin{table}[h]
\caption{
    Outline of default fridge wiring configurations used during systematic optimisation.
}\label{tab:default-configurations}
\begin{tabular*}{\textwidth}{p{3cm}C{1cm}cC{5.2cm}}
\toprule
Line Type&Number&Cable Type&Attenuator Configuration [50K,4K,St,CP,MXC] (dB)\\
\midrule
Drive lines&14&216-SS-SS&[0,20,0,20,20]\\
Flux lines&14&216-CuNi-CuNi&[0,20,0,0,0]\\
Output lines RT-4K&4&119-AgCuNi-CuNi&Not in model\\
Output lines 4K-MXC&4&119-NbTi-NbTi&Not in model\\
\botrule
\end{tabular*}
\end{table}

Throughout this paper, we use a holistic, full-fridge model for systematic optimisations of individual elements of the wiring design, which includes fourteen (14) drive lines, fourteen (14) flux lines and four (4) output lines.
To ensure that the overall heat loads and temperature responses remain approximately representative of the ultimate full-fridge response, and enable fair direct comparisons with established design choices, any lines not being optimised are fixed to default configurations described in Table~\ref{tab:default-configurations}.
Our model incorporates output-line contributions to heat load, but not to device noise.
As this is a small fraction of the total number of lines, contributing an effectively fixed noise background only, this is not expected to impact any of the conclusions drawn from our modelling about any comparisons between different drive-line and flux-line configurations, which was the primary focus of our modelling.
Extending the model to include these effects would be relatively straightforward, apart from requiring characterisation of temperature-dependent input noise generated by low-noise HEMT amplifiers installed at 4K.
(Noise from upper stages is not transmitted below the 4K amplifiers.)
Default drive-line and flux-line attenuator configurations are set to the best-practice solutions identified in~\cite{Krinner2019}.

Throughout this paper, unless otherwise stated, whichever lines are not the subject of a given analysis can be assumed to be set to their default configurations.
To ensure this does not conceal any significant surprising whole-system effects, Fig.~\ref{fig:performance_comparison} then summarises the full-fridge response for our combined optimised solutions, for direct comparison against full existing best-practice solutions.

\section{Temperature Response Modelling of a BlueFors LD-400 fridge}\label{app:temp_response}

\begin{figure}[ht!]
	\centering
 	\includegraphics[width=\linewidth]{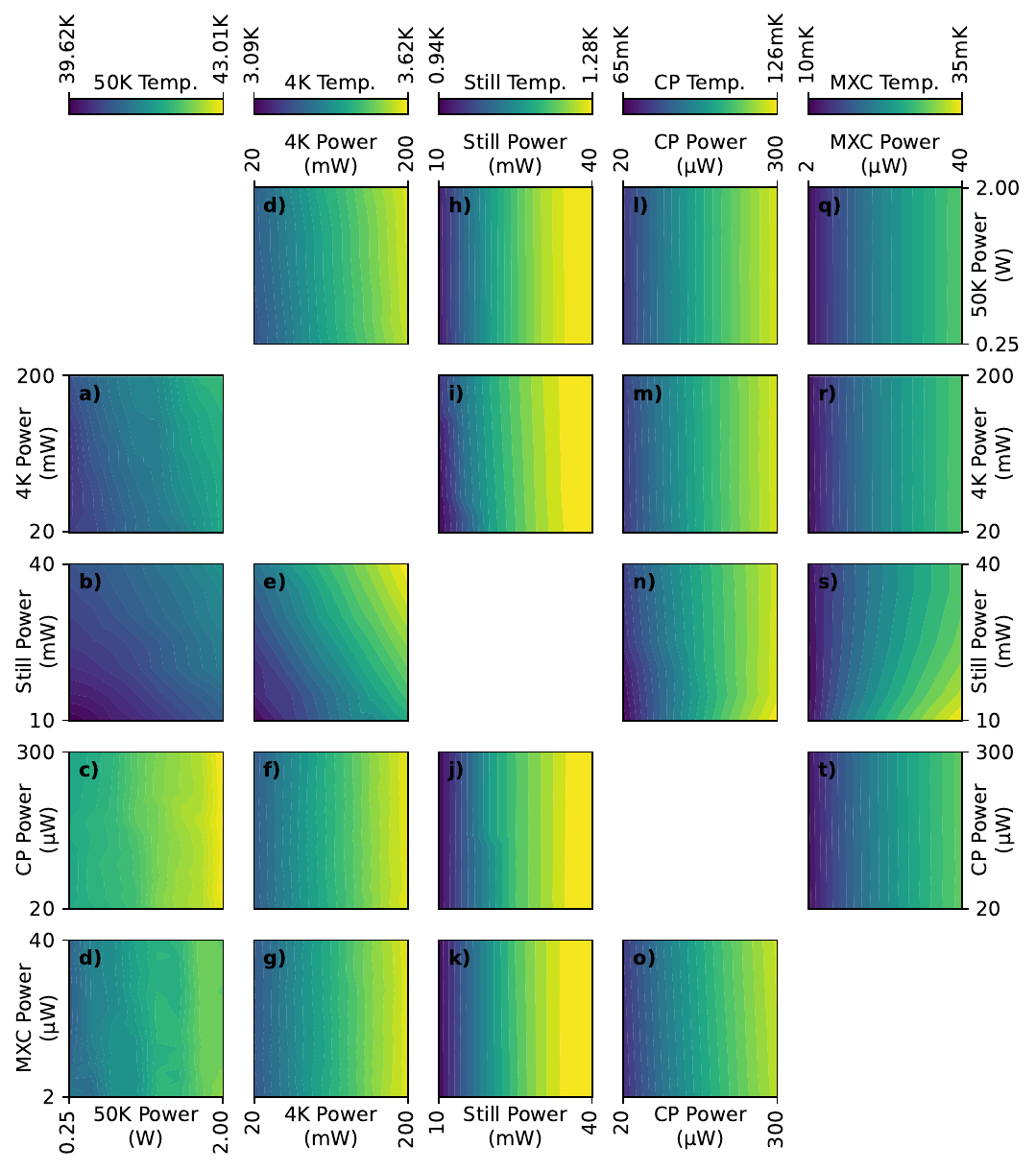}
 	\caption{
        Direct measurements of cross-talk in temperature response between the fridge stages.
        Pairs of applied heat fluxes are swept and the temperature response of all stages is measured.
        Each column gives the temperatures for the 50K (a-d), 4K (e-h), Still (i-l), CP (m-p), and MXC (q-t) stages.
        Apart from interactions with the Still plate, the temperature of each individual plate is dominated by the heat flux directly applied to it.
    }
    \label{fig:crosstalk}
\end{figure}

\begin{figure}[ht!]
	\centering
 	\includegraphics[width=\linewidth]{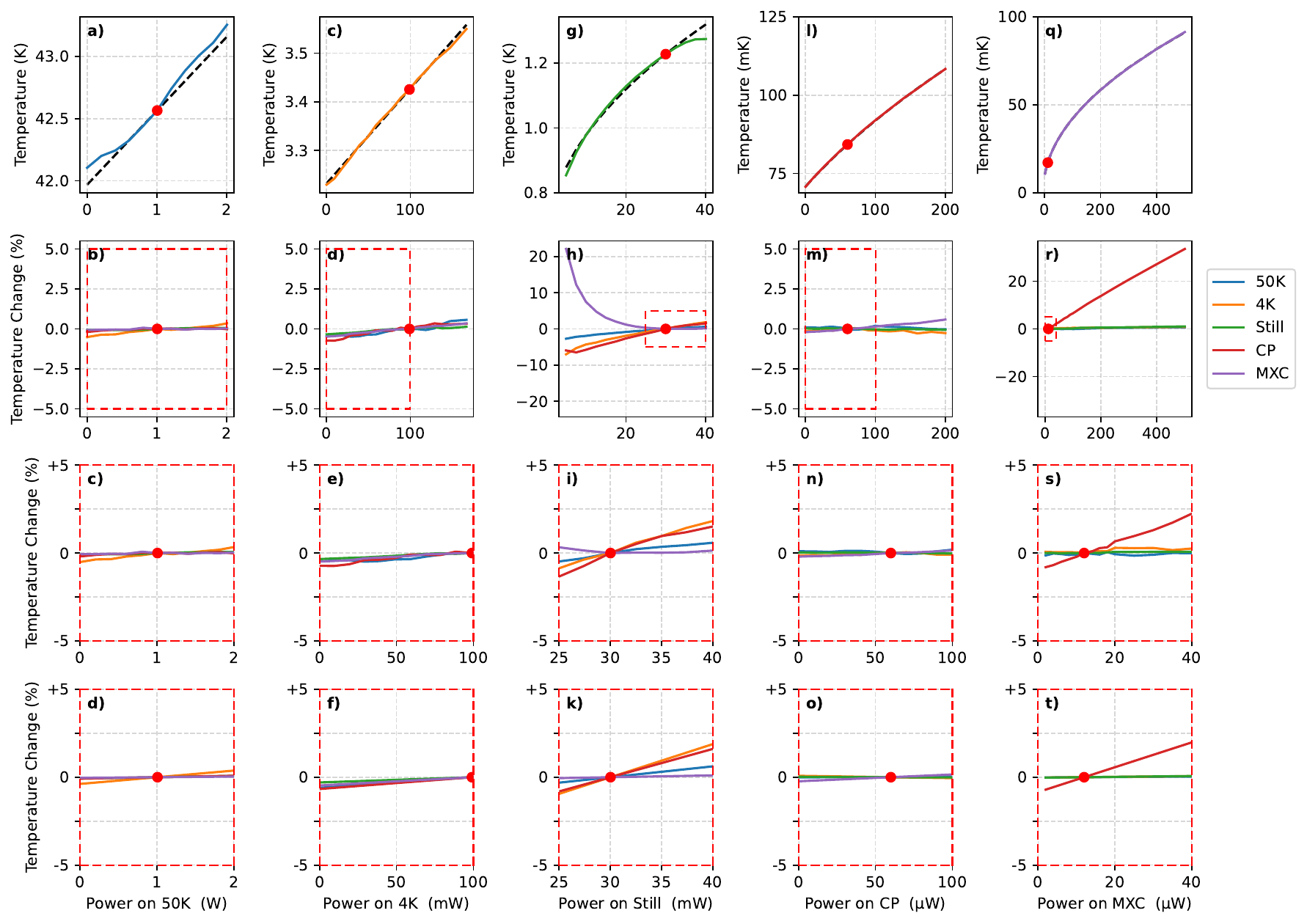}
 	\caption{
        Regression modelling of temperature response of the fridge stages to applied heat fluxes.
        Independent single-variable sweeps of the heat flux are performed on the 50K (a-d), 4K (e-h), Still (i-l), CP (m-p), and MXC (q-t) stages.
        The power applied to all other plates is fixed at the expected heat load power imparted by a full set of cryogenic wiring installed in the fridge.
        The response of each plate to the heat directly applied to is is shown (first row), along with linear or square-root regression fits (black-dashed line).
        The relative change from the set-point temperature of each plate is also shown (second row).
        Changes over the expected heat load power imparted by a full set of cryogenic wiring installed are denoted by red-dashed boxes, and plotted
        separately (third row).
        Linear regression fits (fourth row) accurately model the cross-talk effects over the range of interest, which are small and approximately linear.
    }
    \label{fig:crosstalk2}
\end{figure}

The equilibrium temperature of the dilution refrigerator stages post-installation of cryogenic wiring is a key metric of its performance.
Accurate estimation of the net thermal flux throughout the fridge numerically depends on an intractably large number of physical characteristics
of the system, including the exact geometry, thermal conductivity and emissivity coefficients across the full temperature range of the fridge,
and the countless number of thermal contact resistances at interfaces between components, the structural elements of the fridge and the circulating liquid helium.
Instead, to accurately estimate the expected temperature of the each stage in response to the cryogenic wiring, we can measure the plate responses directly.
We simulate the heat flux channeled or generated by the wiring using electrical electrically controllable heating elements placed at the locations at which the 
cables will be thermalised.
By sweeping the applied heat on each stage and measuring the plate temperatures with 
thermometry, we can map the temperature response in a way which inherently accounts for all of the unique conductances and
radiative loads across the fridge.

Since there is five independent degrees of freedom, namely the heat applied to each of the five stages, mapping the full response space directly is time-consuming
because each measurement requires 10 to 30 minutes to reach thermal equilibrium depending on the plates being driven.
Instead, we aim to fit some regression model to a smaller subset of measurements.
We gather preliminary data, shown in Figure~\ref{fig:crosstalk}, to inform the nature of the regression model, 
importantly, the strength and form of interaction terms between the different stages.
We use heat flux ranges which are indicative of a large range of operating conditions determined using preliminary numerical modelling.
With the exception of the cross-talk and interaction relationships with the Still plate which is a result of this complex and critical role
that the Still plate plays in helium circulation, the temperature of each individual plate is dominated by the heat flux directly applied to it.
The direction of the interpolated isotherms is are mostly linear and have a sharp gradient.
We note that noise from the drift of the 50K plate is likely responsible for the distortion of the temperature isotherms in Figure~\ref{fig:crosstalk}.
The combination of these factors suggests that the interaction terms (such as multiplicative terms) are not significant for the response modelling, 
and the cross-talk effects are small and approximately linear.

On this basis, we perform independent single-variable sweeps of the heat flux applied to each stage and measure the temperature 
response of all stages, shown in Figure~\ref{fig:crosstalk2}.
In each of these sweeps, the power applied to all other plates is fixed at the expected heat load power imparted by a 
full set of cryogenic wiring installed in the fridge, determined by additional preliminary modelling.
This point in heat load space (red point) is referred to as the ``set-point'' and is where all heat load sweeps intersect.
The direct response of each plate to the heat applied follows either a linear relationship or a square-root relationship.
For the MXC plate, the square-root relationship is expected as the cooling power is proportional 
to the square of the plate's temperature: $\dot{Q}_\text{MXC} \propto T_\text{MXC}^2$~\cite{Krinner2019,pobell2007matter}.

We consider the the cross-talk effects as relative changes from the set-point temperature in Figure~\ref{fig:crosstalk2}.
Capturing the cross talk in this way allows us to compare the relative strengths of different pair-wise relationships.
It is also worth noting that there is a ``one-way'' relationship between the upper plates (50K and 4K) and lower plates (CP and MXC)
where the cross-talk coefficients giving the impact of the upper plates on the lower plates is greater than in the reverse direction.
This can be explained considering that the heat load sweeps for the upper plates impart heat inputs into the fridge that 
is more than $10^3$ greater than the expected range of heat loads the lower plates would see directly from cryogenic wiring.

More generally, the strength of the cross-talk relationships is much smaller than the strength of the direct relationships.
If we consider the temperature responses in the operation span of heat load space that we expect to see imparted from
cryogenic wiring, as shown in the third row of Figure~\ref{fig:crosstalk2}, then these cross-talk relationships
are weak. 
By using linear regression models on these relationships to smooth out thermometry noise, 
as shown in the fourth row of Figure~\ref{fig:crosstalk2}, we then have a full and convenient temperature response
model. 
We note, however, that these models do not take into account the change in the effective thermal conductivity between
plates that is introduced by the cryogenic wiring. These effects would change the strength of cross-talk between stages.

We ignore the non-linear cross-talk effects between the Still and other plates, 
as the Still plate is expected to be fixed at a set temperature during operation in order to maximise the cooling power at the device.

\section{Microwave Drive Line attenuation on the 50K}\label{app:50K}

\begin{figure}[ht!]
	\centering
 	\includegraphics[width=\linewidth]{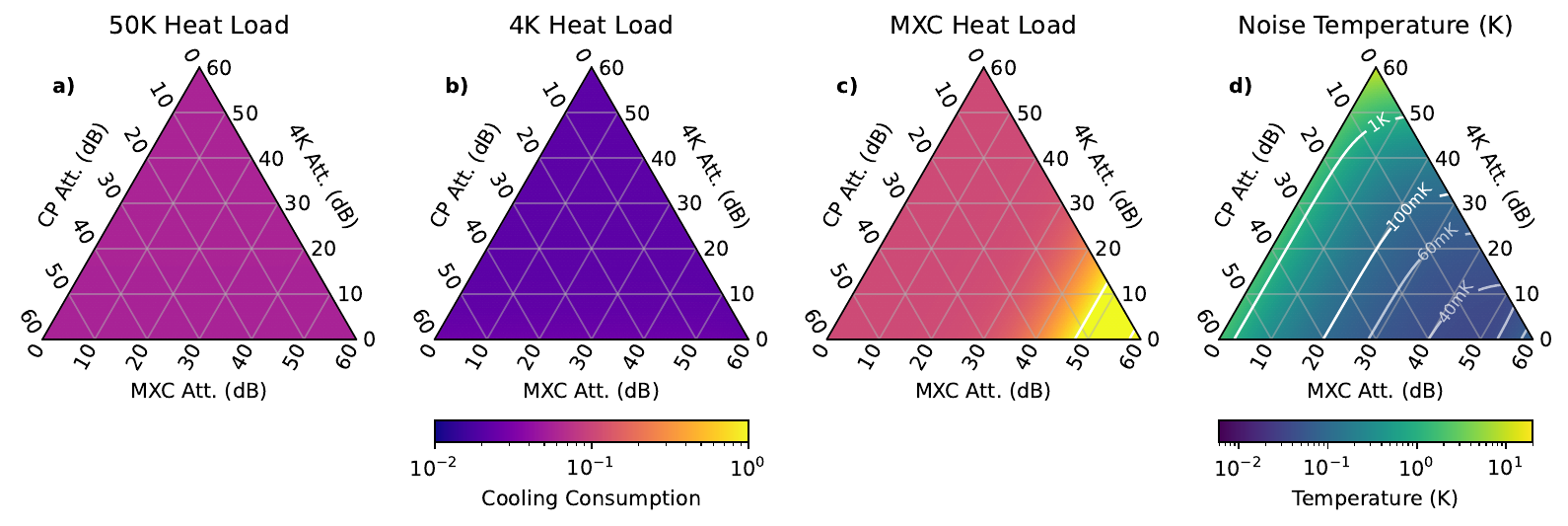}
 	\caption{
        Drive line attenuation configuration space restricted to 50K, 4K, and MXC stages.
        $60\,\text{dB}$ total attenuation is distributed.
        Total system-wide heat loads for the 50K (a), 4K (b), and MXC (c) plates are plotted, along with the resulting drive line noise temperature (e).
        The presence of 50K attenuation reduces the drive line passive heat load by a factor of 4 due to the thermalisation of the inner coaxial conductor (not shown).
        Data is simulated using 14 drive lines, 14 flux lines, and 4 readout lines.
    }
    \label{fig:50K}
\end{figure}

In the same way that using more attenuation above the MXC plate reduces its total cooling consumption, we model
the attenuator configuration space restricted to 50K, 4K, and MXC stages in Figure~\ref{fig:50K} to evaluate if there
is any additional benefit from the inclusion of 50K attenuation on the micrwave drive lines.
The largest benefit is the reduction in passive load cooling consumption on the 4K plateby a factor of 4 due to the 
the thermalisation of the inner conductor at the attenuator on the 50K plate attenuator.
Changing the composition of attenuation, however, does not significantly affect the cooling consumption of the microwave drive 
lines for either of these plates because the passive loads dominate and are independent of the attenuation configuration.

With respect to temperature, lines of constant device and noise temperature are mostly parallel to lines of constant MXC attenuation,
suggesting that only the MXC attenuation provides any meaningful change in noise performance.

Although there is good improvement in the cooling consumption on the 4K plate due to the thermalisation of the inner cable,
only a nominal amount of attenuation is needed to achieve this with no further gains in cooling consumption or noise suppression.
On this basis, we only employ $0\,\text{dB}$ attenuation on the 50K plate, which provides the necessary thermalisation without penalising the noise performance.

\section{Temperature responses of the fridge plates under different total attenuations and contribution ratios}\label{app:new_heuristic}

\begin{figure}[ht!]
	\centering
 	\includegraphics[width=\linewidth]{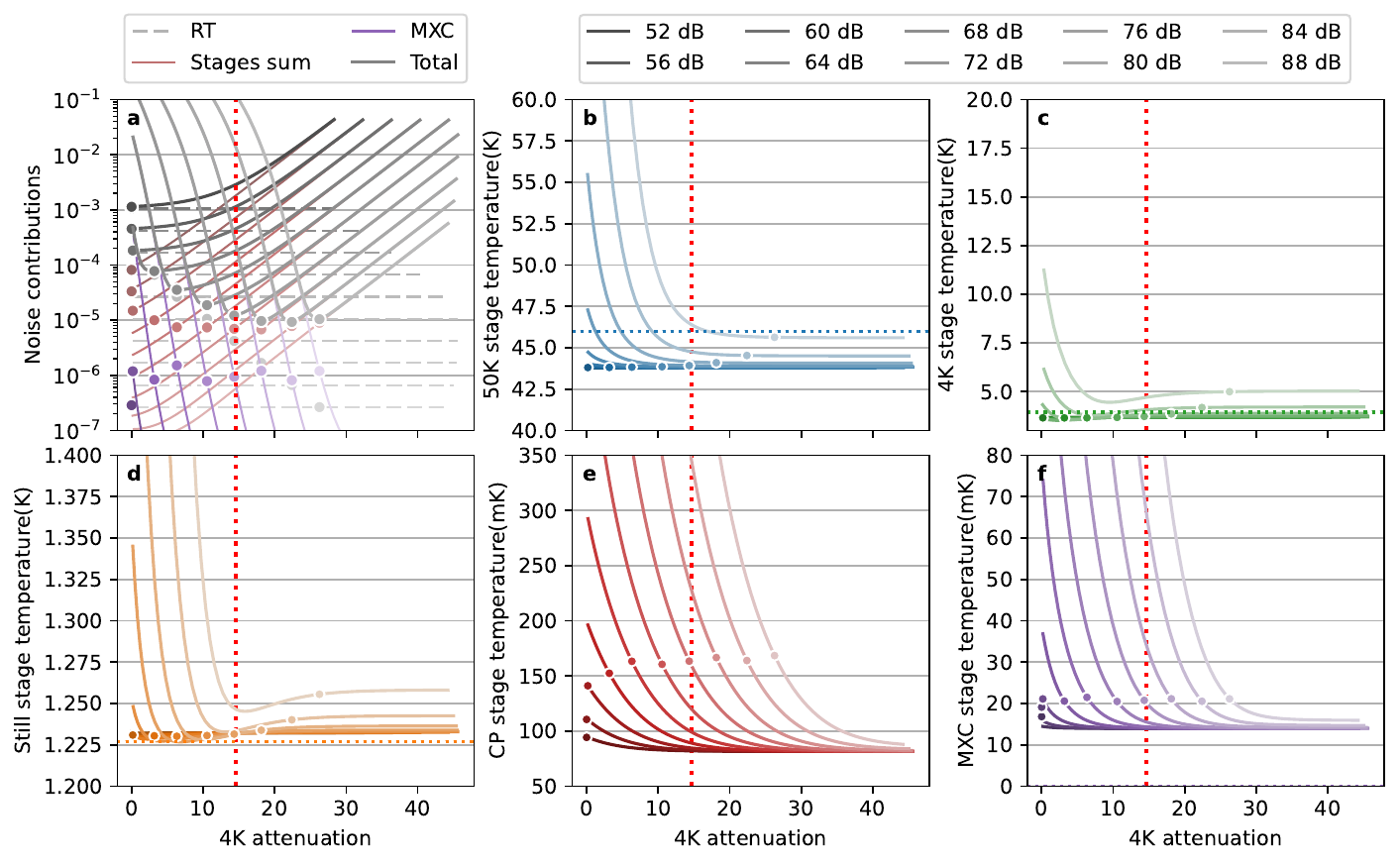}
 	\caption{
        Temperature responses of the fridge plates under different total attenuations and contribution ratios for the drive lines.
        With attenuation restricted to the 4K, CP, and MXC stages, noise contributions are shown (a) for RT, the sum of intermediate stages (4K and CP), and MXC, along with the total noise, for a range of total attenuations.
        As in Figure \ref{fig:drive_line_heuristic}, the attenuation is configured such that the intermediate stage contributions are equal, while the remaining attenuation is put on the MXC stage.
        The corresponding temperature responses for each of these sweeps is shown for the (b) 50K, (c) 4K, (d) Still, (e) CP, and (f) MXC plates.
        Horizontal dotted lines mark the target operating temperatures for each plate.
        The modified ``equal balance'' point for each fixed attenuation where the intermediate stage contributions are equal to the MXC contribution is denoted by filled circles.
        Vertical red dotted lines indicate the point where the noise contributions from the intermediate stages and RT are equal.
    }
    \label{fig:app_fixed_total}
\end{figure}

As discussed in the main text, the temperature response of the cryostat plays an important role in the optimal attenuator configuration of the drive lines which operate in the deep cryogenic regime ($hf\gg k_{\rm B}T$).
Here, we present the temperature responses of the fridge plates under different total attenuations and contribution ratios (controlled by the 4K attenuation) to illustrate the relationship between the temperature responses and the noise contributions from each stage.

The point of equal contributions from the intermediate stages and MXC stage (the modified ``equal balance'' point), denoted by filled circles in Figure~\ref{fig:app_fixed_total}, is a useful reference point since it is where the total noise is minimised for any given total attenuation.
As the total attenuation changes, the value of the CP and MXC temperatures remains roughly constant, while the 4K temperature increases.
This is because to maintain equal contributions from the intermediate stages and MXC stage, only the 4K attenuation can be adjusted to compensate for the change in total attenuation since it effectively modulates only the RT contribution.
This in turn causes only the heat load on the 4K plate to increase and its temperature follows in response.

Below this point for any fixed total attenuation, the temperature of the MXC plate explodes rapidly as the the 4K attenuation is decreased.
This is in response to the resulting increase in its own attenuation and the corresponding increase in its own heat load.
Figure~\ref{fig:app_fixed_total} (a) shows how this temperature response maps to a rapid increase in the MXC contribution to the total noise, which is the dominant factor in the total noise in this region.
In conjunction with the fact that the intermediate stage contributions increase with increasing 4K attenuation, caused by the MXC attenuation decreasing, this why the minimum total noise could necessarily occur at the point: moving in either direction from this point would increase one of the contributions rapidly.

It is important to note that in this modelling, all the plate temperatures eventually explode as the 4K attenuation is decreased.
This is primarily due a result of extrapolation of the temperature cross-talk between the plates in our specific temperature response model as the MXC far exceeds the range of temperatures over which data was collected.
For the purposes of this analysis, however, we are only concerned with the behaviour of the temperature responses in the region around the modified ``equal balance'' point, which is well within the range of our data.

\end{appendices}


\end{document}